\documentclass[a4paper,colorinlistoftodos,prependcaption,textsize=scriptsize]{article}

\usepackage[bottom=1in, left = 1in , right = 1in, top = 1in]{geometry}
\usepackage{setspace}
\usepackage{graphicx}
\usepackage{color} 
\usepackage{enumerate}
\usepackage{fullpage}
\usepackage[pdftex,dvipsnames,table]{xcolor}  
\usepackage{xargs}                      
\definecolor{darkpastelgreen}{rgb}{0.01, 0.75, 0.24}
\definecolor{bostonuniversityred}{rgb}{0.8, 0.0, 0.0}
\definecolor{azzurro}{RGB}{0,122,165}
\usepackage[normalem]{ulem}
\usepackage{soul}
\setstcolor{azzurro}
\sethlcolor{azzurro}

\usepackage[utf8]{inputenc}
\usepackage{mathrsfs,mathtools}
\usepackage{amsmath,amssymb,amsthm,kpfonts,bbm}
\usepackage{amstext}
\usepackage{amsopn}
\usepackage{mathabx}
\usepackage[T1]{fontenc} 
\usepackage{manfnt}
\usepackage{array}
\usepackage{blindtext}
\usepackage{tikz}
\usepackage{easyReview}
\usepackage{float}
\usepackage{multirow}
\usepackage{url}

\usepackage{enumerate}
\usepackage{subcaption}

\newtheorem{proposition}{Proposition}

\newcommand{\llangle}{\left\langle\!\left\langle}
\newcommand{\rrangle}{\right\rangle\!\right\rangle}
\newcommand*\dd{\mathop{}\!\mathrm{d}}
\newcommand{\ie}{{\em i.e.}~}
\newcommand{\eg}{{\em e.g.}~}

\DeclareMathAlphabet{\mathdutchcal}{U}{dutchcal}{m}{n}
\usepackage{physics} 
\def \C {\boldsymbol{\mathdutchcal{C}}}
\def \S {\boldsymbol{\mathdutchcal{S}}}
\newcommand{\sgn}[1]{\mathrm{sign}(#1)}

\title{Periodically forced pinned anharmonic atom chains}
\author{S.~Darshan$^{1}$, A.~Iacobucci$^{2}$, S.~Olla$^{3,4,5}$ and G.~Stoltz$^{6,7}$ \\
	\small $^1$ ESSEC Business School, Cergy\\
	\small $^2$ CNRS \& CEREMADE, Universit\'e Paris-Daphine, PSL University, 75016 Paris, France\\
	\small $^3$ CEREMADE, Universit\'e Paris-Daphine, PSL University, 75016 Paris, France\\
    \small $^4$ Institut Universitaire de France \\
    \small $^5$ GSSI, L'Aquila, Italy\\
    \small $^6$ CERMICS, ENPC, Institut Polytechnique de Paris, Marne--la--Vall\'ee\\
    \small $^7$ MATHERIALS team-project, Inria Paris, France\\
}

\begin{document}
	
	\maketitle

\begin{abstract}
	Recent works proved a hydrodynamic limit for periodically forced atom chains with harmonic interaction and pinning, together with momentum flip \cite{KLO1,KLO2}. When energy is the only conserved quantity, one would expect similar results in the anharmonic case, as conjectured for the temperature profile and energy flux in \cite{Presutti_bday}. However, outside the harmonic case, explicit computations are generally no longer possible, thus making a rigorous proof of this hydrodynamic limit difficult. Consequently, we numerically investigate the plausibility of this limit for the particular case of a chain with \(\beta\)--FPUT interactions and  harmonic pinning. We present our simulation results suggesting that the PDE for the limiting temperature profile and the Green--Kubo type formula for the limiting energy current conjectured in \cite{Presutti_bday} are correct. We then use this Green--Kubo type formula to investigate the relationship between the energy current and period of the forcing. This relationship is investigated in the case of significant rate of momentum flip, small rate of momentum flip and no momentum flip. We compare the relationship observed in the anharmonic case to that of the harmonic case for which explicit formulae are available \cite{KLO1}.
\end{abstract}

\section{Introduction}

Atom chains have served since the early 1960s as fundamental models for studying energy transport in low-dimensional systems, see for example \cite{LLP, Dhar08, LLP16}. A long-standing open question is whether such systems satisfy Fourier’s law, that is, whether they exhibit normal diffusion characterized by a well-defined, finite thermal conductivity~$\kappa$ independent of the size of the system. Purely harmonic chains violate Fourier’s law due to their ballistic transport; their conductivity scales linearly with the length of the system $\kappa \sim n$, see \cite{RLL, Nakazawa1, Nakazawa2}. In acoustic chains, i.e those with total momentum conservation, the introduction of nonlinear interactions typically leads to anomalous transport, with $\kappa \sim n^\alpha$ for some $0 < \alpha < 1$. An exception is provided by rotor chains, which display normal conductivity despite their momentum conservation (see~\cite{GS00}).
To recover diffusive behavior ($\alpha = 0$), total momentum conservation must be broken, either by the presence of an on-site (pinning) potential or through stochastic perturbations acting in the bulk.

In this work, we are interested in the conversion of mechanical energy into heat, and in the behavior of the resulting energy current. We consider a harmonically pinned chain of~$n$ atoms with unit mass, interacting through an anharmonic $\beta$--FPUT potential. The system is coupled to a Langevin--type heat bath at the left boundary and subjected to a time--periodic forcing at the right boundary. Such a model was initially considered in \cite{KLO1,KLO2} with harmonic pinning and interaction potentials as well as momentum flip and  in~\cite{Presutti_bday,PBS} with anharmonic pinning and harmonic interaction potentials without momentum flip. However, in order to obtain a well defined large size limit, it is in fact necessary to suitably rescale the intensity of the forcing by inverse of the square root of the system size~$n$, i.e. \(\frac{1}{\sqrt{n}}\). Indeed, numerical simulations suggest that without this rescaling the temperature of the forced atom explodes as the system size increases, see \cite[Section IV.C]{KMKD}. In our model, energy is the only conserved quantity. The presence of the on-site potential breaks translational invariance, thereby destroying both momentum and stretch conservation. In addition, the Hamiltonian dynamics in the bulk is perturbed by a momentum--flip noise: Each particle randomly flips the sign of its momentum at exponentially distributed times. This noise guarantees a provably finite thermal conductivity \cite{Cedric_Stefano} and introduces an additional dissipative mechanism that contributes to the conversion of mechanical energy, supplied by the external forcing, into thermal energy absorbed by the thermostat. Furthermore, momentum–flip noise tends to suppress finite-size effects, so the limit properties become apparent at computationally tractable system sizes. In this work, we are less concerned with finite-size effects, which have already been investigated in detail for a closely related model in \cite{PBS, KMKD}. The periodic boundary forcing continuously injects energy into the system and ensures the emergence of a time-periodic stationary state, with the same period as the applied force, sustaining a non-zero average energy current across the chain. Our main objective is to numerically investigate the plausibility of the hydrodynamic limit for this system.

Recently, the linear (harmonic) version of the above model has been rigorously analyzed in \cite{KLO1,KLO2}. Among the main results, the authors establish: 
\begin{enumerate}[(i)]
	\item the existence and uniqueness of a time-periodic stationary state, absolutely continuous with respect to Lebesgue measure, to which the system converges at large times from any initial configuration;
	\item an explicit expression for the one-period averaged work done by the external force, valid for all system sizes;
	\item the conditions on the scaling of the force amplitude and its period with respect to the system size~$n$ that yield a spatially constant average energy current in the thermodynamic limit $n \to \infty$;
	\item an explicit expression for the thermal conductivity~$\kappa$ in the linear response regime, which is shown to be independent of temperature;
	\item a rigorous derivation of the macroscopic heat equation satisfied by the temperature profile, obtained through a diffusive space-time scaling limit.
\end{enumerate}
The limiting partial differential equation (PDE) features a Dirichlet boundary condition on the thermostatted end and a Neumann-type condition at the driven boundary and admits a linear temperature profile as stationary solution.
The unpinned case has been rigorously studied in~\cite{KOS24}, where the authors consider a periodic forcing containing multiple frequencies. In this setting, both energy and volume stretch are conserved, and their macroscopic evolution is governed, in the diffusive limit, by a system of coupled partial differential equations. The energy profile satisfies a modified heat equation with a source term depending on the gradient of the stretch field, which accounts for the conversion of mechanical energy into heat. The authors also establish a clear distinction between the mechanical and thermal contributions to the total work performed by the periodic force, showing that it splits into a mechanical part, carried by the low-frequency modes and associated with coherent deformation of the system, and a thermal part, linked to the high-frequency components and responsible for irreversible dissipation.
These results were further extended in~\cite{Presutti_bday}, where the authors study a variant of the model with thermostats at both boundaries and periodic forcing applied at a site in the bulk. In this setting, the total energy current is divided into mechanical and thermal components whose interplay governs the transport properties of the system. The temperature profile has a slope discontinuity at the forcing site and, in the unpinned case, the conservation of volume stretch leads to a parabolic profile.

A variant of the harmonically pinned linear model has been studied in~\cite{Garrido}, where the chain is coupled to Langevin thermostats at both boundaries and periodically forced at the rightmost site, without any bulk noise (i.e., no momentum flips). The system reaches a Gaussian, time-periodic stationary state in the long-time limit. However, due to the absence of bulk noise and the lack of scaling of the force with system size (see~\cite{Presutti_bday}), it is not expected that the dynamics will admit a hydrodynamic limit. The work and mechanical energy depend sensitively on whether the forcing frequency is in the harmonic spectrum. When the frequency lies outside the spectrum, both quantities vanish in the thermodynamic limit, while for frequencies within the spectrum, they exhibit fast oscillations.

We present simulation results for $\beta$-FPUT chains, supporting both the conjectured partial differential equation for the macroscopic temperature profile and the Green--Kubo-type formula for the limiting energy current conjectured in~\cite{Presutti_bday}. 
Our numerical investigation validates the hydrodynamic PDE through two complementary approaches: first, by comparing the temperature profiles obtained from direct simulations with the solution of the PDE, using transport coefficients and boundary data extracted from simulations of the microscopic system; and second, by confirming that the Green–Kubo-type estimate of the average energy flux matches the observed one. We also study how the forcing period affects the energy current and the work performed by the external force at different momentum-flip rates.
The numerical results indicate that energy transmission persists for forcing frequencies above the harmonic band, even as the flip intensity decreases, a phenomenon known as supratransmission~\cite{KLR}. Remarkably, this effect remains present even in the absence of momentum flips, in contrast to the purely harmonic case~\cite{Garrido} and the case of interactions with a bounded anharmonic part studied by perturbation analysis in \cite{Garrido2}. In fact in \cite{Garrido2}, it is proved rigourously that with bounded small anharmonicity and no random flip in the dynamics, no supratransmission occurs: outside the harmonic band the work done converges to 0 exponentially fast with the size of the system. This suggests a conjecture that supratransmission effects are due to the unboundness of the anharmonicity, such as the presence of the quartic term in the $\beta$-FPUT interaction. Moreover, the upper edge of the transmission band increases with temperature, highlighting the role of anharmonicity in extending energy propagation beyond the linear regime. Finally, as the flip rate is reduced, the response of the system exhibits a pronounced peak near the lower edge of the harmonic spectrum, suggesting the onset of resonance-like behavior in the anharmonic chain.

The remainder of this paper is organized as follows. In Section~\ref{sec:theory} we define the microscopic model and describe the expected macroscopic behavior in the hydrodynamic limit, including the conjectured partial differential equation for the stationary temperature profile. Section~\ref{sec:numerics} briefly describes the numerical methods used to simulate the dynamics and compute observables such as temperature, current, and work, and presents the results obtained. We compare the numerically obtained temperature profiles with the solution of the hydrodynamic equation and test the validity of the Green–Kubo-type formula for the average energy current. Moreover, we study frequency-dependent phenomena, including supratransmission and the emergence of resonance-like behavior. We conclude in Section~\ref{sec:conclusions} with a brief discussion of the results and perspectives for future work. The appendices provide the full details of the numerical methods used.

\section{Model and Expected Behavior in the Hydrodynamic Limit} \label{sec:theory}
In Section~\ref{sec:model}, we present the microscopic dynamics of our model. In Section~\ref{sec:limite_hydro}, we heuristically derive a non--linear partial differential equation with a non--linear Neumann boundary condition that describes the expected stationary temperature profile in the limit \(n \to \infty\). We also give in Section~\ref{sec:gk_work_deriv} a heuristic derivation of the Green--Kubo-like formula for the rate of work done by the periodic forcing in the hydrodynamic limit conjectured in \cite{Presutti_bday}. We then give in Section~\ref{sec:exc_uni_sol_pde} sufficient conditions for the existence and uniqueness of a solution to the limiting PDE.

\subsection{Presentation of the model}\label{sec:model}
Our microscopic model consists of a chain of \(n+1\) atoms pinned via a potential \(u\)
and interacting with their nearest neighbors via a potential \(v\). We suppose that the interaction potential has its global minimum at \(0\) and that \(v(0) = 0\).  This hypothesis can be made to hold by translating \(v\) and adding a constant to it. In the numerical simulations in Section~\ref{sec:numerics}, we will consider the $\beta$-FPUT interaction potential 
$v(r) = \frac{r^2}{2} + \frac{r^4}{4}$ and harmonic pinning potential $u(q) = \frac{k_0 q^2}{2}$. In this section, we consider general potentials under the assumption that the dynamics and corresponding Gibbs measure are well defined.

We add a thermostat on the leftmost particle at temperature \(T_\ell\) with intensity~\(\gamma\), and drive the rightmost particle with a periodic forcing of period \(\theta\). The non-thermostatted particles, \ie all but the first one, are subject to random momentum flip with rate \(\widetilde{\gamma}\).
We denote the particle positions by \(q = \left(q_0, \dots, q_n\right) \in \mathbb{R}^{n+1}\) and their momenta by \(p = \left(p_0, \dots, p_n\right) \in \mathbb{R}^{n+1}\). The dynamics of the chain reads
\begin{equation}\label{eq:dynamics}
	\begin{aligned}
		\dd q_x &= p_x \, \dd t, && x \in \left\{0, 1, \dots, n\right\},\\
		\dd p_0 &= \left(v'\left(q_1 - q_0\right) - u'\left(q_0\right)\right)\dd t -\gamma p_0 \, \dd t + \sqrt{2\gamma T_\ell} \,\dd W_t,\\
		\dd p_x &= \left(v'\left(q_{x+1} - q_x\right) - v'\left(q_x - q_{x-1}\right) - u'\left(q_x\right)\right)\dd t - 2p_x\left(t-\right)\dd N_x\left(\widetilde{\gamma}t\right), && x \in \left\{1, \dots, n-1\right\},\\
		\dd p_n &= \left(-v'\left(q_n - q_{n-1}\right) - u'\left(q_n\right)\right)\dd t - 2p_n\left(t-\right)\dd N_n\left(\widetilde{\gamma}t\right) + \frac{1}{\sqrt{n}}\mathdutchcal{F}\left(\frac{t}{\theta}\right)\dd t,
	\end{aligned}
\end{equation}
where \(W\) is a one-dimensional standard Brownian motion and \(\left\{N_x\right\}_{x = 1}^n\) is a collection of independent rate 1 Poisson jump processes independent of \(W\). Here \(\mathdutchcal{F}\) is a 1-periodic function. By rescaling time by \(\widetilde{\gamma}\), the Poisson jump process \(\left(N_x\left(\widetilde{\gamma} t\right)\right)_{t\geq 0}\) becomes a process that makes unit jumps with rate \(\widetilde{\gamma}\) and \(\dd N_x\left(\widetilde{\gamma}t\right)\) corresponds to a unit impulse given at the jump times of said Poisson process. So the \(-2p_x\left(t-\right)\dd N_x\left(\widetilde{\gamma}t\right)\) term subtracts from \(p_x\) twice its value at each jump time, which amounts to flipping the sign of \(p_x\) whenever an exponential clock with rate \(\widetilde{\gamma}\) rings.
The generator of the dynamics \eqref{eq:dynamics} is
\begin{equation}\label{eq:generator}
	\mathdutchcal{G}_t = \mathdutchcal{L}_0 + \frac{1}{\sqrt{n}}\widetilde{\mathdutchcal{L}}_t,
\end{equation}
with \(\mathdutchcal{L}_0 = \mathdutchcal{A} + \widetilde{\gamma}\mathdutchcal{S}_{\mathrm{flip}} + \gamma \mathdutchcal{S}_{0, T_\ell}\) the generator of the equilibrium dynamics and \(\widetilde{\mathdutchcal{L}}_t = \mathdutchcal{F}\left(\frac{t}{\theta}\right)\partial_{p_n}\) the generator of the periodic forcing. In the expression of the generator of the equilibrium dynamics, \(\mathdutchcal{A}\) is the Liouville operator given by
\[\mathdutchcal{A} = \sum_{x = 0}^n \left[p_x \partial_{q_x} + \left(v'\left(q_{x+1} - q_x\right) - v'\left(q_x - q_{x-1}\right) - u'\left(q_x\right)\right)\partial_{p_x}\right],\]
where we use the convention that \(q_{-1} \equiv q_0\) and \(q_{n+1} \equiv q_n\).
As for the momentum flip operator \(\mathdutchcal{S}_{\mathrm{flip}}\), given a function \(f : \mathbb{R}^{n+1} \times \mathbb{R}^{n+1} \to \mathbb{R}\), it acts as
\[\mathdutchcal{S}_{\mathrm{flip}}f(q, p) = \sum_{x = 1}^n \left[f(q, p^x) - f(q, p)\right],\]
where \(p^x\) is the momentum vector with the \(x\)-th coordinates sign changed, i.e. \(p_y^x = p_y\) for \(y \neq x\) and \(p^x_x = -p_x\). Lastly, \(\mathdutchcal{S}_{0, T_\ell}\) is the generator of the thermostat given by
\[\mathdutchcal{S}_{0, T_\ell} = -p_0\partial_{p_0} + T_\ell\partial_{p_0}^2.\]
Adopting the same convention \(q_{-1} \equiv q_0\) and \(q_{n+1} \equiv q_n\), we define the on-site energy as 
\[\mathdutchcal{E}_x(q,p) = \frac{p_x^2}{2} + u\left(q_x\right) + v\left(q_x - q_{x-1}\right), \qquad x \in \left\{0, \dots, n\right\},\]
so that the Hamiltonian of the system is given by
\begin{equation}
	\mathdutchcal{H}(q,p) = \sum_{x = 0}^n \mathdutchcal{E}_x(q,p) = \sum_{x=0}^n\left(\frac{p_x^2}{2} + u\left(q_x\right) + v\left(q_x - q_{x-1}\right)\right).
\end{equation}
The dynamics~\eqref{eq:dynamics} locally conserves the energy, indeed
\[\mathdutchcal{G}_t\mathdutchcal{E}_x = j_{x-1, x} - j_{x, x+1},\qquad i \in \{1,\dots,n-1\}\]
where $j_{x, x+1}$ is the local energy flux, which is given in the bulk by
\begin{equation}\label{eq:bulk_flux_def}
	j_{x, x+1} = -p_x v'\left(q_{x+1} - q_x\right), \qquad x \in \left\{0, \dots, n-1\right\},
\end{equation}
and at the boundaries by
\begin{equation}\label{eq:bd_flux_def}
	j_{-1, 0} = \gamma\left(T_\ell - p_0^2\right), \qquad j_{n, n+1} = -\frac{1}{\sqrt{n}}\mathdutchcal{F}\left(\frac{t}{\theta}\right)p_n.
\end{equation}
Note that the right boundary flux $j_{n, n+1}$ is the opposite of the instantaneous work done by the forcing. This is in line with the usual convention that a positive flux corresponds to energy moving from left to right and a positive work corresponds to energy being injected into the system. 

We denote by \(\mu_T^n\) the invariant probability measure of the equilibrium dynamics at temperature \(T\), i.e. the dynamics of the unforced system in contact with the thermostat at temperature \(T\). This measure is given by the Boltzmann--Gibbs distribution
\begin{equation}
	\mu_T^n\left(\dd q\,\dd p\right) \propto \exp\left(-\frac{1}{T}\mathdutchcal{H}(q, p)\right)\dd q\,\dd p.
\end{equation}
We suppose that the dynamics \eqref{eq:dynamics} admits an invariant \(\theta\)--periodic cycle, \ie a unique \(\sigma\)-finite measure \(\rho_n\) on \(\mathbb{R}\times\mathbb{R}^{n+1}\times\mathbb{R}^{n+1}\) that is invariant under time translations of length \(\theta\):
\[\int_0^{\infty}\int_{\mathbb{R}^{n+1}\times\mathbb{R}^{n+1}} A(t,q,p)\rho_n\left(\dd t\, \dd q\, \dd p\right) = \int_{k\theta}^{\infty}\int_{\mathbb{R}^{n+1}\times\mathbb{R}^{n+1}} A(t-k\theta,q,p)\rho_n\left(\dd t\, \dd q\, \dd p\right), \qquad \forall k \in \mathbb{N}\]
and such that \(\mathbf{1}_{[0,\theta]}(t)\rho_n\left(\dd t\,\dd q\, \dd p\right)\) is a probability measure. We also denote by \(\llangle\cdot \rrangle_{n}\) the integral with respect to \(\rho_n\) over one period for the time variable and the phase space for the position and momentum variables, i.e. for a test function \(A\),
\[\llangle A \rrangle_{n} = \int_0^\theta\int_{\mathbb{R}^{n+1} \times \mathbb{R}^{n+1}} A(t, q, p)\rho_n\left(\dd t\, \dd q\, \dd p\right).\]
For a smooth observable \(A\) that are \(\theta\)--periodic in time, we have \(\llangle \mathdutchcal{G}_t A \rrangle_{n} = 0\). This holds in particular for the energy at each site: \(\llangle \mathdutchcal{G}_t \mathdutchcal{E}_x  \rrangle_{n} = 0\). The local conservation of energy implies that the energy flux is uniform along the chain at the steady state, i.e.
\begin{equation}\label{eq:stat_flux}
	\forall x \in \left\{0, \dots, n-1\right\}, \qquad \llangle j_{x, x+1} \rrangle_{n} = \llangle j_{-1, 0}\rrangle_{n} = \llangle j_{n, n+1}\rrangle_{n},
\end{equation}
thus the total average energy flux is 
\[J_n := \sum_{x=0}^{n-1} \llangle j_{x, x+1}\rrangle_{n}= n \,\llangle j_{0, 1} \rrangle_{n}.\]
We define the temperature profile on the basis of microscopic dynamics as
\begin{equation}
	\begin{aligned}
		T_{\mathrm{ss}}^n(u) &= \llangle p_x^2\rrangle_{n}, \qquad && \text{for } u \in \left[\frac{x}{n}, \frac{x+1}{n}\right), \quad x\in \left\{0, \dots, n-1\right\},\\
		T_{\mathrm{ss}}^n(1) &= \llangle p_n^2\rrangle_{n}.
	\end{aligned}
\end{equation}

\subsection{Expected behavior in the hydrodynamic limit}\label{sec:limite_hydro}
Fourier's law predicts that the energy density flux \(j\) is proportional to the opposite of the temperature gradient, that is at steady state
	\begin{equation}\label{eq:fourier_law}
		j(u) = -D\left(T_{\mathrm{ss}}(u)\right) \nabla T_{\mathrm{ss}}(u),
	\end{equation}
where \(D(T)\) is the thermal conductivity, whose expression is made precise below in \eqref{eq:thermalcond}. Since the microscopic energy flux is constant along the chain at the steady state \eqref{eq:stat_flux}, one expects the energy density flux to be constant in the hydrodynamic limit as well \(j(u,t) = J = \lim_{n \to\infty}J_n\).

The microscopic counterpart of~\eqref{eq:fourier_law} is
\[J_n \, \propto - n\left(T_{\mathrm{ss}}^n\left(\frac{x+1}{n}\right) - T_{\mathrm{ss}}^n\left(\frac{x}{n}\right)\right),\qquad \forall x \in \{0,\dots,n-1\},\]

If a hydrodynamic limit holds, \(T_{\mathrm{ss}}^n\) converges to some limiting profile \(T_{\mathrm{ss}}\) and \(J_n\) converges to some finite constant value \(J\). Consequently, taking the limit as \(n\) goes to infinity in the above expression, we recover \eqref{eq:fourier_law}. The limiting stationary temperature profile \(T_{\mathrm{ss}}\) therefore satisfies 
\begin{equation}\label{eq:bulk_pde}
	\partial_u \left(D\left(T_{\mathrm{ss}}(u)\right) \partial_u T_{\mathrm{ss}}(u)\right) = 0, \qquad \forall u \in [0,1],
\end{equation}
with boundary conditions depending on those imposed on the microscopic dynamics.

In our case, the thermostat on the left boundary corresponds to a Dirichlet boundary condition at \(u = 0\):
\[T_{\mathrm{ss}}(0) = T_\ell,\]
while the periodic driving corresponds to a non-linear Neumann boundary condition at \(u = 1\): 
\[D\left(T_{\mathrm{ss}}(1)\right)\partial_uT_{\mathrm{ss}}(1) = -J = \mathbb{W}\left(T_{\mathrm{ss}}(1), \mathdutchcal{F}, \theta\right),\]
where \(\mathbb{W}\left(T, \mathdutchcal{F}, \theta\right)\) is the rate of work done by the $\theta$-periodic force \(\mathdutchcal{F}\) at temperature \(T\). The last equality follows from the fact that the periodic forcing is the only mechanism driving the system out of equilibrium and generating a net flux. Consequently, in the steady state the energy flux \(J\) across the system is the opposite of the work rate $W$ done by the forcing, given our sign conventions: \(J>0\) denotes energy transport from left to right, while \(W>0\) denotes work done on the system (i.e. energy added to it). 

In \cite{Presutti_bday}, the authors give the following expression for~\(\mathbb{W}\), which can be derived formally via a linear response argument (see Appendix~\ref{sec:gk_work_deriv}):
\begin{equation}\label{eq:gk_work}
	\mathbb{W}\left(T, \mathdutchcal{F},\theta\right) = \frac{1}{T}\int_0^\infty \left(\frac{1}{\theta}\int_0^\theta \mathdutchcal{F}\left(\frac{s}{\theta}\right) \mathdutchcal{F}\left(\frac{s+t}{\theta}\right)\dd s\right)\mathbb{E}_T^+\left[p_0(t)p_0(0)\right]\dd t,
\end{equation}
where \(\mathbb{E}_T^+\) is the expectation with respect to the semi-infinite equilibrium dynamics started under the Boltzmann--Gibbs measure with temperature \(T\). Precisely, the dynamics \(\left(q_x, p_x\right)_{x \in \mathbb{N}}\) in \eqref{eq:gk_work} satisfies
\begin{equation}\label{eq:semi_infinite_eq_chain}
	\begin{aligned}
		\dd q_x &= p_x \, \dd t, && x \in \mathbb{N},\\
		\dd p_0 &= \left(v'\left(q_1 - q_0\right) - u'\left(q_0\right)\right)\dd t -2p_0\dd N_0\left(\widetilde{\gamma}t\right),\\
		\dd p_x &= \left(v'\left(q_{x+1} - q_x\right) - v'\left(q_x - q_{x-1}\right) - u'\left(q_x\right)\right)\dd t - 2p_x\dd N_x\left(\widetilde{\gamma}t\right), && x \in \mathbb{N}\setminus \left\{0\right\},
	\end{aligned}
\end{equation}
and the Boltzmann--Gibbs measure is formally given by
\begin{equation}\label{eq:semi_infinite_boltzmann-gibbs}
	\mu_T^+\left(\dd q \, \dd p\right) \propto \prod_{\phantom{+}x \in \mathbb{N}} \exp\left(-\frac{1}{T}\left\{\frac{p_x^2}{2} + v\left(q_x - q_{x-1}\right) + u\left(q_x\right)\right\}\right)\dd q_x \, \dd p_x.
\end{equation}
We refer to \cite[Theorem 2.3]{FFL} for the technical details on the well-posedness of the dynamics and the definition of the Boltzmann--Gibbs measure of the semi--infinite system.

Putting this together, we expect that in the hydrodynamic limit the temperature profile satisfies the following transport equation
\begin{equation}\label{eq:limite_hydro}
	\left\{\begin{aligned}
		&\partial_u \left[D\left(T_{\mathrm{ss}}(u)\right) \partial_u T_{\mathrm{ss}}(u)\right] = 0,\\
		&T_{\mathrm{ss}}(0) = T_\ell,\\
		&D\left(T_{\mathrm{ss}}(1)\right)\partial_uT_{\mathrm{ss}}(1) = \mathbb{W}\left(T_{\mathrm{ss}}(1), \mathdutchcal{F}, \theta\right),
	\end{aligned}\right.
\end{equation}
where the diffusion coefficient $D(T)$ is the thermal conductivity which appears in~\eqref{eq:fourier_law} and is given by (see \emph{e.g.} \cite{AlessandraStefanoGabriel})
\begin{equation}\label{eq:thermalcond}
	\begin{aligned}
		D(T) &= \lim_{n\to \infty}\frac {1}{n T^2} \int_0^{+\infty} \sum_{\phantom{+}x = 0}^{n-1} \sum_{y = 0}^{n-1} \, \mathbb{E}_T^+[j_{y,y+1}(0)\,j_{x,x+1}(t)] \, \dd t \\
		&= \lim_{n\to \infty}\frac {1}{nT^2} \int_0^{+\infty} \, \mathbb{E}_T^+\left[\left(\sum_{y = 0}^{n-1} j_{y,y+1}(0)\right)\left(\sum_{x = 0}^{n-1} j_{x,x+1}(t)\right)\right] \, \dd t.
	\end{aligned}
\end{equation}
Sufficient conditions for the existence of solutions of \eqref{eq:limite_hydro} and a characterization of the set of these solutions are given in Section~\ref{sec:exc_uni_sol_pde}

\subsection{Existence of a solution to \eqref{eq:limite_hydro}}\label{sec:exc_uni_sol_pde}
In this subsection, we use Schauder's fixed-point theorem to show that \eqref{eq:limite_hydro} admits a solution under some hypotheses on \(D\) and \(\mathbb{W}\). We say that a function \(T_{\mathrm{ss}} \in C^1\left([0,1], \mathbb{R}_+\right)\) is a mild solution of \eqref{eq:limite_hydro} if it satisfies the boundary conditions 
\[T_{\mathrm{ss}}(0) = T_\ell, \qquad D\left(T_{\mathrm{ss}}(1)\right)\partial_u T_{\mathrm{ss}}(1) = \mathbb{W}\left(T_{\mathrm{ss}}(1), \mathdutchcal{F}, \theta\right),\] 
and
\begin{equation}\label{eq:fixed_point_problem}
	T_{\mathrm{ss}}(u) = T_\ell + \mathbb{W}\left(T_{\mathrm{ss}}(1), \mathdutchcal{F}, \theta\right) \int_0^u \frac{\dd v}{D\left(T_{\mathrm{ss}}(v)\right)}.
\end{equation}
\begin{proposition}\label{prop:mild_sol}
	Let \(T_\ell \geq 0 \). Suppose that the conductivity \(D\) is continuous and uniformly bounded below by a positive constant (i.e there exists \(\varepsilon > 0\) such that \(D(T) \geq \varepsilon\) for all \(T \geq 0\)) and that for a fixed forcing~\(\mathdutchcal{F}\) and period \(\theta\), \(\mathbb{W}\) is a function of the temperature, which grows at most sublinearly, \ie there exists \(K > 0\) and \(\alpha \in [0, 1)\) such that,
	\[\forall T \geq 0, \qquad \mathbb{W}\left(T, \mathdutchcal{F}, \theta\right) \leq K\left(1 + T^\alpha \right).\]
	Then, the equation \eqref{eq:limite_hydro} admits a mild solution. 
\end{proposition}

We believe that the conductivity \(D\) and the work rate \(\mathbb{W}\) satisfy the above two hypotheses in the case we study numerically, \ie \(\beta\)--FPUT interactions, harmonic pinning, and cosine forcing. Indeed, according to our simulations, these hypotheses appear to hold. See Figures~\ref{fig:cond_fit}~and~\ref{fig:work_fit} in Appendix~\ref{sec:correlation} for some plots of a representative set of parameters that substantiate this belief.\nocite{KipnisLandimOlla}

\begin{proof}
	The integral equation \eqref{eq:fixed_point_problem} defines a map \(\mathscr{M}: C\left([0,1], \mathbb{R}_+\right) \to C\left([0,1], \mathbb{R}_+\right)\) given by
	\begin{equation}\label{eq:M}
		\mathscr{M}[f](u) = T_\ell + \mathbb{W}\left(f(1), \mathdutchcal{F}, \theta\right) \int_0^u\frac{\dd v}{D\left(f(v)\right)}.
	\end{equation}
	for \(f \in C\left([0, 1], \mathbb{R}_+\right)\).
	If \(T_{\mathrm{ss}}\) is a fixed point of \(\mathscr{M}\), then \(T_{\mathrm{ss}}\) satisfies the boundary conditions of \eqref{eq:limite_hydro} and \(T_{\mathrm{ss}} \in C^1\left([0,1], \mathbb{R}\right)\) since
	\[\partial_u T_{\mathrm{ss}}(u) = \partial_u \mathscr{M}\left[T_{\mathrm{ss}}\right](u) = \frac{\mathbb{W}\left(T_{\mathrm{ss}}(1), \mathdutchcal{F},\theta\right)}{D\left(T_{\mathrm{ss}}(u)\right)}.\]
	The rightmost term in the previous identity is continuous since \(D\) is continuous and uniformly bounded below. Consequently, \(T_{\mathrm{ss}}\) is a mild solution of \eqref{eq:limite_hydro}.
	
	For \(M > 0\), denote by \(\mathdutchcal{B}_M := \left\{g \in C\left([0,1], \mathbb{R}_+\right) \left| \, \|g\|_\infty := \sup_{u \in [0,1]} g(u) \leq M \right.\right\}\) the closed ball of radius \(M\) in \(C\left([0,1], \mathbb{R}\right)\) intersected with \(C\left([0,1], \mathbb{R}_+\right)\). We will show that, for \(M\) large enough, \(\mathscr{M}\) is well defined from \(\mathdutchcal{B}_M\) to \(\mathdutchcal{B}_M\) and that \(\mathscr{M}\) is a compact operator. Then, since \(\mathdutchcal{B}_M\) is a nonempty bounded closed convex subset of the Banach space \(C\left([0,1], \mathbb{R}\right)\), Schauder's fixed-point theorem implies that \(\mathscr{M}\) has a fixed point in \(\mathdutchcal{B}_M\); see for example \cite[Theorem 2.A]{Zeidler}. 
	
	Consider \(f \in C\left([0,1], \mathbb{R}_+\right)\) such that \(\|f\|_\infty \leq M\). Then
	\[\sup_{u \in [0, 1]} \mathscr{M}[f](u) \leq T_\ell + \mathbb{W}\left(f(1), \mathdutchcal{F}, \theta\right)\sup_{u \in [0, 1]}\int_0^u\frac{\dd v}{D\left(f(v)\right)} \leq T_\ell   +  \frac{K\left(1 + M^\alpha\right)}{\varepsilon}.\]
	For \(M > 0\) large enough, 
	\[T_\ell + \frac{K\left(1 + M^\alpha\right)}{\varepsilon} \leq M, \]
	so \(\left\|\mathscr{M}[f]\right\|_\infty \leq M\), thus \(\mathscr{M}\) maps \(\mathdutchcal{B}_M\) into itself. 
	
	We now estimate \(\left\|\partial_u\mathscr{M}[f]\right\|_\infty\) for \(f \in C\left([0,1], \mathbb{R}_+\right)\) such that \(\|f\|_\infty \leq M\):
	\[\sup_{u \in [0, 1]} \left|\partial_u \mathscr{M}[f](u)\right| \leq \mathbb{W}\left(f(1), \mathdutchcal{F}, \theta\right) \frac{1}{D\left(f(u)\right)} \leq \frac{K}{\varepsilon}\left(1 + M^\alpha\right).\]
	Consequently, for \(M > 0\),
	\[\mathscr{M}\left[\mathdutchcal{B}_M\right] \subset \left\{f \in C^1\left([0, 1], \mathbb{R}_+\right) \left|\, \|f\|_\infty \leq \left(T_\ell + \frac{K}{\varepsilon}\left(1 + M^\alpha\right)\right)\!\!, \,\, \|\partial_u f\|_\infty \leq \frac{K}{\varepsilon}\left(1 + M^\alpha\right) \right. \right\}.\]
	By the Arzelà--Ascoli theorem, the set on the right-hand side of the previous inclusion relation is relatively compact in \(C\left([0,1], \mathbb{R}\right)\) and thus so is \(\mathscr{M}\left[\mathdutchcal{B}_M\right]\). Therefore, \(\mathscr{M}\) is a compact operator. Having satisfied the hypotheses of Schauder's fixed-point theorem, we can conclude that \(\mathscr{M}\) admits a fixed point in \(\mathdutchcal{B}_M\). 
\end{proof}

\paragraph{(Non-)Uniqueness of the Solution to \eqref{eq:limite_hydro}.} By Proposition~\ref{prop:mild_sol}, the PDE \eqref{eq:limite_hydro} admits at least one mild solution. We define the following function
\[A(x) = \int_0^x D(y)\dd y.\]
The next proposition characterizes the set of mild solutions to \eqref{eq:limite_hydro}.
\begin{proposition}
	Suppose that the conductivity \(D\) is continuous and uniformly bounded below by a positive constant. Fix the temperature on the left hand side \(T_\ell \geq 0\), the forcing \(\mathdutchcal{F}\), and the forcing period \(\theta\). Then \eqref{eq:limite_hydro} admits as many solutions as the scalar equation 
	\begin{equation}\label{eq:scalar_eq}
		A\left(T^\star\right) = A\left(T_\ell\right) + \mathbb{W}\left(T^\star, \mathdutchcal{F}, \theta\right).
	\end{equation}
\end{proposition}

One condition that ensures that \eqref{eq:scalar_eq} admits at most one solution is that the function \(T \mapsto A\left(T\right) - \mathbb{W}\left(T, \mathdutchcal{F}, \theta\right)\) is strictly monotone. We believe that this function is strictly increasing for the case of \(\beta\)-FPUT interactions, harmonic pinning, and cosine forcing. Our belief is reinforced by the observation that this property appears to hold in all of our simulations. We include in Appendix~\ref{sec:numerical_verification_strict_increasing} some plots of the numerically estimated values \(A - \mathbb{W}\) for a representative set of parameters we simulated.

\begin{proof}
	Suppose that \(T^\star\) is a solution of \eqref{eq:scalar_eq} and define for, \(u \in [0,1]\), the function
	\begin{equation}\label{eq:def_T_candidate}
		T(u) = A^{-1}\left(A\left(T_\ell\right) + \mathbb{W}\left(T^\star, \mathdutchcal{F}, \theta\right)u\right).
	\end{equation}
	The inverse of \(A\) is well defined since it is a strictly increasing function as its derivative \(D\) is strictly positive. Furthermore, since \(D\) is continuous, \(A \in C^1\left(\mathbb{R}_+, \mathbb{R}_+\right)\) and so is \(T\) by the inverse function theorem in one dimension. Deriving \(T\), we get
	\[\partial_u T (u) = \frac{\mathbb{W}\left(T^\star, \mathdutchcal{F}, \theta\right)}{D\left(A^{-1}\left(A\left(T_\ell\right) + \mathbb{W}\left(T^\star, \mathdutchcal{F}, \theta\right)u\right)\right)} = \frac{\mathbb{W}\left(T^\star, \mathdutchcal{F}, \theta\right)}{D\left(T(u)\right)},\]
	where the second equality comes from recognizing the expression of \(T\) in the argument of \(D\). Integrating both sides and rearranging gives
	\[T(u) = T(0) + \mathbb{W}\left(T^\star, \mathdutchcal{F}, \theta\right) \int_0^u \frac{\dd v}{D\left(T(v)\right)}.\]
	From the definition \eqref{eq:def_T_candidate} of \(T\), we have \(T(0) = T_\ell\) and furthermore the definition of \(T\) and \eqref{eq:scalar_eq} imply that \(T(1) = T^\star\). Consequently, \(T\) is a mild solution of \eqref{eq:limite_hydro}.
	
	Now suppose that \(T\) is a mild solution of \eqref{eq:limite_hydro}. The composition of two \(C^1\) function \(A\circ T\) is also~\(C^1\) and its derivative is
	\[\partial_u A\left(T(u)\right) = D\left(T(u)\right) \partial_u T(u) = \mathbb{W}\left(T(1), \mathdutchcal{F}, \theta\right),\]
	where the second equality comes from the fact that, as a mild solution of \eqref{eq:limite_hydro}, \(T\) satisfies \eqref{eq:fixed_point_problem}. Integrating the far left and right sides of the above equality, we obtain
	\[A\left(T(u)\right) - A\left(T(0)\right) = \mathbb{W}\left(T(1), \mathdutchcal{F}, \theta\right)u.\]
	It is immediate from \eqref{eq:fixed_point_problem} that \(T(0) = T_\ell\). Evaluating the above equality at \(u = 1\) and rearranging gives
	\[A\left(T(1)\right) = A\left(T_\ell\right) + \mathbb{W}\left(T(1), \mathdutchcal{F}, \theta\right),\]
	i.e. \(T(1)\) solves \eqref{eq:scalar_eq}.
\end{proof}

\section{Numerical Results}\label{sec:numerics}
We present in this section some of the results of our numerical investigations of periodically forced chains with \(\beta\)-FPUT interactions \(v(r) = \frac{r^4}{4} + \frac{r^2}{2}\) and harmonic pinning \(u(q) = \frac{k_0 q^2}{2}\), with $k_0 = 1$. In Section~\ref{sec:setup}, we briefly describe the numerical schemes used to simulate the microscopic dynamics and estimate the rate of work done by the periodic forcing. Further details are deferred to Appendices~\ref{sec:correlation}~\&~\ref{sec:solving_pde}. In Section~\ref{sec:order_1_flip}, we present the simulation results for \(\widetilde{\gamma} = 1\). In Section~\ref{sec:small_flip_rate}, we present the simulation results for small to vanishing flip rates. We pay particular attention to the differences between the small-flip rate regime and the order 1 flip rate regime. We conclude in Section~\ref{sec:no_flip} by presenting the results of simulations without momentum flip.

\subsection{Numerical setup}\label{sec:setup}
Unless otherwise specified, all of our simulations\footnote{The code for these simulations is available at \url{github.com/shiva-darshan/beta-fput-under-periodic-forcing}} are performed with a thermostat with friction parameter \(\gamma = 1\) and a chain of length \(n + 1 = 2000\). Preliminary simulations indicate that, in the presence of momentum-flip noise, a length of \(n+1 = 2000\) is sufficient to observe the chain’s asymptotic behavior. We use a cosine driving
\[\mathdutchcal{F}\left(\frac{t}{\theta}\right) = f_0 \cos\left(\frac{2\pi}{\theta} t\right), \]
where \(f_0\) is the forcing intensity and \(\theta\) the forcing period. We report our simulation results as a function of the forcing intensity and frequency
\begin{equation}\label{eq:freq_convention}
    \nu = \frac{1}{\theta}.
\end{equation}
In each of the following subsections, we first specify the forcing parameters (intensity and frequency) used in the simulations, and then present the corresponding results.
\paragraph{Integration of dynamics.}
In all cases, the dynamics is integrated with a BAOAB scheme \cite{BDM} with a time step of \(\Delta t = 0.005\) and momentum flips during the O-step. The numerical scheme is the following:
\begin{equation}\label{eq:discrete_dynamics}
	\left\{\begin{aligned}
		p_0^{k+1/3} &= p_0^k + \left(v'\left(q_1^{k} - q_0^{k}\right) - u'\left(q_0^k\right)\right)\Delta t/2,\\
		p_x^{k+1/3} &= p_x^k + \left(v'\left(q_{x+1}^{k} -q_x^{k} \right) - v'\left(q_x^{k} - q_{x-1}^{k}\right) - u'\left(q_x^{k}\right)\right)\Delta t/2, && x \in \left\{1, \dots, n-1\right\},\\
		p_n^{k+1/3} &= p_n^k +\left(-v'\left(q_n^k - q_{n-1}^k\right) - u'\left(q_n^k\right) + \frac{1}{\sqrt{n}}\mathdutchcal{F}\left(\frac{\left(k+1/2\right)\Delta t}{\theta}\right) \right)\Delta t/2,\\
		q^{k+1/2}_x &= q^k_x +  p_x^{k+1/3} \Delta t/2, && x \in \left\{0, 1, \dots, n\right\},\\
		p_0^{k + 2/3} &= \mathrm{e}^{-\gamma \Delta t}p_0^{k+1/3} + \sqrt{T_\ell \left(1 - \mathrm{e}^{-2\gamma\Delta t}\right)}G^{k+1},\\
		p_x^{k+2/3} &= p_x^{k+1/3} - 2p_x^{k+1/3}\mathdutchcal{B}_x^k, && x\in \left\{1, \dots, n\right\},\\
		q^{k+1} &= q^{k+1/2}_x + p^{k+2/3}_x \Delta t/2, && x\in \left\{0, 1, \dots, n\right\}\\
		p_0^{k+1} &= p_0^{k+2/3} + \left(v'\left(q_1^{k+1} - q_0^{k+1}\right) - u'\left(q_0^{k+1}\right)\right)\Delta t/2,\\
		p_x^{k+1} &= p_x^{k +2/3}  + \left(v'\left(q_{x+1}^{k+1} -q_x^{k + 1} \right) - v'\left(q_x^{k+1} - q_{x-1}^{k+1}\right) - u'\left(q_x^{k + 1}\right)\right)\Delta t/2, && x \in \left\{1, \dots, n-1\right\},\\
		p_n^{k+1} &= p_n^{k+2/3} +\left(-v'\left(q_n^{k+1} - q_{n-1}^{k+1}\right) - u'\left(q_n^{k+1}\right) + \frac{1}{\sqrt{n}}\mathdutchcal{F}\left(\frac{\left(k+1\right)\Delta t}{\theta}\right) \right)\Delta t/2,\\
	\end{aligned}
	\right .
\end{equation}
where \(\left\{G^k\right\}_{k \in\mathbb{N}}\) are independent and identically distributed (i.i.d.) real-valued standard Gaussian random variables and \(\left\{\mathdutchcal{B}_x^k\right\}_{k \in \mathbb{N}, \, x \in \left\{1, \dots, n\right\}}\) are  i.i.d Bernoulli random variables with \(\mathbb{P}\left(\mathdutchcal{B}_x^k = 1\right) = \frac{1}{2}\left(1 - \mathrm{e}^{-2\widetilde{\gamma}\Delta t}\right)\).
This probability is the probability of the Poisson process \(N_x\left(\widetilde{\gamma}t\right)\) making an odd number of jumps in a time interval of length \(\Delta t\), i.e. the sign of the momentum of the \(x\)-th particle changing over a time interval of length \(\Delta t\). 
Observe that in the middle step the momentum flip and the Ornstein--Uhlenbeck update at site $x=0$ commute, as they act on different particles. 
This scheme is equivalent to a palindromic scheme and thus should be of weak order two in the absence of forcing \cite{LMS, BDM}.
We fix the temperature of the thermostat on the left to \(T_\ell = 0.1\) for all the simulations.

\paragraph{Computation of temperature profiles and bulk energy fluxes.}
To compute the temperature profile and bulk flux, we first initialize the atom chain at \(\mu_{T_\ell}^n\). The initialization is done by starting the chain with all positions equal to zero and momenta sampled independently from the Gaussian distribution \(\mathcal{N} (0,T_\ell)\) by adding thermostats at temperature \(T_\ell\) to each site and running the dynamics for \(10^6\) time steps.
We then burn-in the dynamics~\eqref{eq:discrete_dynamics} for \(2\times 10^9\) time steps, and finally run it for \(N = 10^{10}\) time steps to compute time averages.

For the computation of temperature profiles, we estimate \(\llangle p_x^2\rrangle_{\mathrm{ss}, n}\) at each site \(x \in \left\{0, 1, \dots, n\right\}\) with a time average over the simulated trajectory
\begin{equation}\label{eq:temperature_estimator}
	\llangle p_x^2\rrangle_{\mathrm{ss}, n} \approx \frac{1}{N}\sum_{k = 1}^N \left(p_x^k\right)^2 - \left(\frac{1}{N}\sum_{k=1}^N p_x^k\right)^2.
\end{equation}
We subtract the empirical momentum average to attenuate the spurious effect of time discretization. To estimate error bars on temperature profiles, we cut the trajectory into 100 equal-sized batches (subtrajectories) of length \(N_{\mathrm{batch}} = 10^{8}\) and compute the variance of the empirical averages among the batches as follows
\[\mathrm{Var} \approx \frac{1}{B - 1}\sum_{b = 1}^{B} \left(\frac{1}{N_{\mathrm{batch}}}\sum_{k = 1}^{N_{\mathrm{batch}}}\left(p_x^{k + bN_{\mathrm{batch}}}\right)^2 - \left(\frac{1}{N_{\mathrm{batch}}}\sum_{k=1}^{N_{\mathrm{batch}}} p_x^{k + bN_{\mathrm{batch}}}\right)^2 - \llangle p_x^2\rrangle_{\mathrm{ss}, n}\right)^2,\]
with \(B = 100\) the number of batches.
Since the batches are sufficiently long, we observe a negligible correlation among them and consider them as uncorrelated.
To estimate the total average energy flux \(J_n = n\llangle j_{0, 1}\rrangle_{\mathrm{ss}, n}\), we exploit the fact that, at stationarity, it should be constant in the bulk. Consequently, we take the time average over the simulated trajectory and the spatial average over the bulk sites
\begin{equation}\label{eq:bulk_flux_estimator}
	J_n \approx  \frac{1}{N} \sum_{x = 0}^{n-1} \sum_{k=1}^N j_{x, x+1}^k, \qquad j_{x, x+1}^k = -p_x^k v'\left(q_{x+1}^k - q_x^k\right), \quad x \in \left\{0, \dots, n-1\right\}.
\end{equation}

We compare the solutions~\(T(u)\) of the conjectured limiting PDE \eqref{eq:limite_hydro} to the empirical profiles observed via direct simulation. With the same simulation runs used to compute the autocorrelation of the momentum of the first particle, we compute the autocorrelation of the energy flux at equilibrium to estimate the conductivity \(D(T)\) at various temperatures, see Appendix~\ref{sec:correlation} for details. We can then view \(T \mapsto D(T)\) and \(T \mapsto \mathbb{W}^n\left(T, f_0, \theta\right)\) (see \eqref{eq:finite_size_W} below) as functions of the temperature (in the case of \(\mathbb{W}^n\), a function which is parameterized by the forcing parameters \(f_0\) and \(\theta\)). We fit the functions \(T \mapsto D(T)\) and \(T \mapsto \mathbb{W}^n\left(T, f_0, \theta\right)\) to the data obtained by simulation. With these approximations, we solve the PDE \eqref{eq:bulk_pde} by solving a discretized fixed point problem detailed in Appendix~\ref{sec:solving_pde} (which is a numerical counterpart to the fixed point problem introduced in the proof of Proposition~\ref{prop:mild_sol}).

\paragraph{Estimation of the rate of work.}
To estimate the rate of work \eqref{eq:gk_work} at temperature \(T\), we replace the semi-infinite dynamics with the dynamics of a finite chain \eqref{eq:finite_eq_chain} initialized at \(\mu_T^n\). With our choice of forcing, we can explicitly compute the autocorrelation of \(\mathdutchcal{F}\) as a function of \(t\),
\[\frac{1}{\theta}\int_{0}^\theta \mathdutchcal{F}\left(\frac{t+s}{\theta}\right)\mathdutchcal{F}\left(\frac{s}{\theta}\right)\dd s = \frac{f_0^2}{2}\cos\left(\frac{2\pi t}{\theta}\right). \]
Our finite $n$ approximation of \eqref{eq:gk_work} is then given by
\begin{equation}\label{eq:finite_size_W}
	\mathbb{W}^n\left(T, f_0, \theta\right) = \frac{f_0^2}{2T} \int_0^\infty \cos\left(\frac{2\pi}{\theta}t\right)\mathbb{E}_{\mu_T^n}\left[p_0(t)p_0(0)\right]\dd t. 
\end{equation}
To numerically compute this integral, we split it into two parts:
\begin{equation}\label{eq:split_W_integrals}
	\mathbb{W}^n\left(T, f_0, \theta\right) = \frac{f_0^2}{2T} \int_0^{t_0} \cos\left(\frac{2\pi}{\theta}t\right)\mathbb{E}_{\mu_T^n}\left[p_0(t)p_0(0)\right]\dd t + \frac{f_0^2}{2T} \int_{t_0}^\infty \cos\left(\frac{2\pi}{\theta}t\right)\mathbb{E}_{\mu_T^n}\left[p_0(t)p_0(0)\right]\dd t.
\end{equation}
The computation of the first-particle momentum autocorrelation is made precise in Appendix~\ref{sec:correlation}. The first integral is then computed by direct quadrature via the trapezoidal rule using the numerically estimated momentum autocorrelation function:
\[\begin{aligned}
	&\int_0^{t_0} \cos\left(\frac{2\pi}{\theta}t\right)\mathbb{E}_{\mu_T^n}\left[p_0(t)p_0(0)\right]\dd t \\
	&\quad\approx \Delta t\sum_{k = 1}^{\left \lfloor \frac{t_0}{\Delta t}\right\rfloor - 1} \cos\left(\frac{2\pi \Delta t k}{\theta}\right) \widehat{\mathtt{C}}_p[k] +  \frac{\Delta t}{2}\left(\widehat{\mathtt{C}}_p[0] + \cos\left(\frac{2\pi \Delta t \left\lfloor t_0/\Delta t\right\rfloor}{\theta}\right) \widehat{\mathtt{C}}_p\left[\left\lfloor t_0/\Delta t\right\rfloor\right]\right),
\end{aligned}\]
where \(\widehat{\mathtt{C}}_p\) is the empirical autocorrelation function estimated from simulation, see \eqref{eq:num_p_corr} in Appendix~\ref{sec:correlation}. 
For the second integral, we fit a function \(\widehat{C}(t)\) to the tail of the autocorrelation of the momentum, 
\[\mathbb{E}_{\mu_T^n}\left[p_0(t)p_0(0)\right] \approx \widehat{C}(t), \qquad t \geq t_0. \]
We tested several functional forms and identified the following as particularly effective. We introduce the parameters \(K>0\), \(\lambda>0\), \(\nu^\star >0\) and \(\phi\in \left[0, 2\pi\right]\). For \(\widetilde{\gamma}\) of order 1, we use
\begin{equation}\label{eq:Cp1}
	\widehat{C}_{p,1}\left(t; K, \lambda \right) = K \mathrm{e}^{-\lambda t};
\end{equation}
for \(0 < \widetilde{\gamma} < 1\) not too small, we use
\begin{equation}\label{eq:Cp2}
	\widehat{C}_{p,2}\left(t; K, \lambda, \nu^\star, \phi\right) = K\mathrm{e}^{-\lambda t}\cos\left(2\pi\nu^\star t + \phi\right);
\end{equation}
for \(0 < \widetilde{\gamma} \ll 1 \), we use
\begin{equation}\label{eq:Cp3}
	\widehat{C}_{p,3}\left(t; K, \lambda, \nu^\star, \phi\right) = K\frac{\mathrm{e}^{-\lambda t}\cos\left(2\pi\nu^\star t + \phi\right)}{\sqrt{t}};
\end{equation}
finally, for \(\widetilde{\gamma} = 0\), we use
\begin{equation}\label{eq:Cp4}
	\widehat{C}_{p,4}\left(t; K, \nu^\star, \phi\right) = K \frac{\cos\left(2\pi\nu^\star t + \phi\right)}{\sqrt{t}}, \qquad \nu^\star \neq \nu.
\end{equation}
We found that the exponential decay rate of the tail of the correlation function is of the same order as the momentum flip rate, namely $\lambda = \mathdutchcal{O}(\widetilde{\gamma})$. When the flip rate is very small, and hence the exponential decay is slow, the medium--time behavior of the correlation function is dominated by a slower polynomial decay, which we attribute to the Hamiltonian part of the dynamics. As a consequence, for sufficiently small flip rates, the approximation must include a factor $t^{-1/2}$ to correctly capture the tail behavior.
These functional forms allow the analytical computation of the second integral in \eqref{eq:split_W_integrals} as a function of the forcing and the fitted parameters (see Appendix~\ref{sec:tail_integrals} for the formulae we use for computing the second integral in \eqref{eq:split_W_integrals} in each regime of the flip rate and their derivations).
We compare the work rate of the case $\beta$ -FPUT with that of the harmonic case, given in the hydrodynamic limit by the following explicit formula derived in \cite[Theorem 3.1]{KLO1}:
\[
\mathbb{W}^{\rm harm} = - 4\gamma \, \left(\frac{2\pi}{\theta}\right)^2 \sum_{\ell \in \mathbb{Z}} \ell^2 \left| \widetilde{\mathdutchcal{F}}(\ell)\right|^2 \int_0^1 \cos^2 \left(\frac{\pi z}{2}\right) \left\{\left[ 4 \sin^2 \left(\frac{\pi z}{2}\right) + 1 - \left(\frac{2\pi\ell}{\theta}\right)^2\right]^2 + \left(\frac{4\gamma\pi\ell}{\theta}\right)^2\right\}^{-1}\dd z,
\]
where \(\widetilde{\mathdutchcal{F}}(\ell) = \int_0^1 {\rm e}^{-2\pi \mathrm{i} \ell t} \mathdutchcal{F}(t) \dd t\) is the Fourier transform of the external periodic force $\mathdutchcal F$. Note that $\mathbb{W}^{\rm harm}$ is independent of the temperature.

\paragraph{The harmonic band.}
The eigenvalues of \(-\Delta_n\), the discrete Laplacian with Neumann boundary conditions are
\[
    \lambda_j = 4 \sin^2\left(\frac{\pi j}{2(n+1)}\right), \qquad j\in\{0,\dots, n\},
\]
see, for example, \cite[Section 2.3]{KLO1}. Therefore, the harmonic band in the pinned case is the half--open interval that contains the square roots of the eigenvalues of \(k_0 - \Delta_n\), 
\[\omega_j = \sqrt{k_0 + 4 \sin^2 \left(\frac{\pi j}{2(n+1)}\right)}, \qquad j \in \left\{0, \dots, n\right\},\]
namely
\(\left[\sqrt{k_0}, \sqrt{k_0 + 4}\right)\),
where \(k_0\) is the strength of the harmonic pinning potential \(u\).
In view of our convention for the frequencies \eqref{eq:freq_convention}, we divide by \(2\pi\) to obtain what we use as the harmonic band of frequencies
\begin{equation}
\nu \in \left[\frac{\sqrt{k_0}}{2\pi}, \frac1{2\pi} \sqrt{k_0 + 4}\,\right) = \left[\frac{1}{2\pi}, \frac{\sqrt 5}{2\pi}\right)\approx [0.159, 0.356),
\end{equation}
We set to \(k_0 = 1\) throughout and denote by $\widebar{\nu} = \frac1{2\pi}$ the lower bound of the harmonic band.

\subsection{Moderate/large flip rate \(\widetilde{\gamma} = 1\)}\label{sec:order_1_flip}
In this subsection we report the results of simulations with momentum flip rate \(\widetilde{\gamma} = 1\). We start by comparing the empirical energy flux \eqref{eq:bulk_flux_estimator} of the dynamics \eqref{eq:dynamics} to the numerical estimate of the rate of work \eqref{eq:finite_size_W}. We next consider the relationship between the forcing frequency and the rate of work. We finish the section by comparing the temperature profiles observed in direct simulation to those obtained by solving the PDE \eqref{eq:limite_hydro}.

\begin{figure}[h]
	\centering
	\includegraphics[width = 0.7\textwidth]{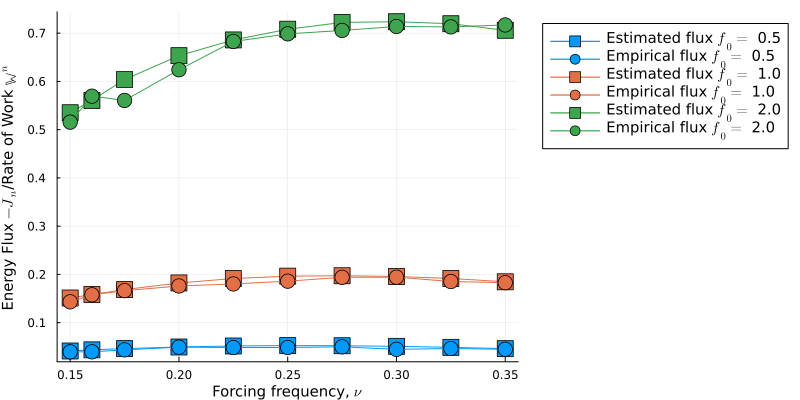}
	\caption{Empirical negative flux observed in simulations (circles) and rate of work estimated using \eqref{eq:finite_size_W} (squares), both plotted against frequency within the harmonic band, using the same forcing parameters and evaluated at the temperature of the forced atom site.}
	\label{fig:emp_flux_v_work_flip1}
\end{figure}

\paragraph{Validity of the Green--Kubo like formula.}
In Figure~\ref{fig:emp_flux_v_work_flip1}, we plot the values of the empirical bulk energy flux (or rather its opposite) \(-J_n\) and the estimated value of rate of work \(\mathbb{W}^n\) at the temperature empirically observed at the forced site, see Table~\ref{tab:flip1} in Appendix~\ref{sec:correlation} for the values of this temperature. We consider the following values of forcing frequencies and intensities:
\[\left(\nu, f_0\right) \in \left\{0.15, 0.16, 0.175, 0.2, 0.225, 0.25, 0.275, 0.3, 0.325, 0.35\right\} \times \left\{0.5, 1.0, 2.0\right\}.\]
There is good agreement between the empirically observed flux \eqref{eq:bulk_flux_estimator} and the numerically computed rate of work~\eqref{eq:finite_size_W} over the explored range of forcing parameters. This agreement suggests that the expression of \(\mathbb{W}^n\) given by \eqref{eq:finite_size_W} accurately estimates the limiting energy flux/work rate.

\begin{figure}[h]
	\centering
	\includegraphics[width = 0.6\textwidth]{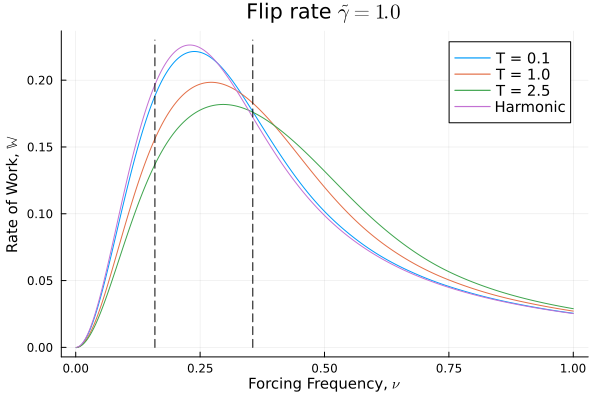}
	\caption{Work rate estimated using \eqref{eq:finite_size_W} plotted against the forcing frequency \(\nu\), with \(f_0 = 1\). The harmonic band 
	\(\nu \in \left[\widebar{\nu}, \frac{\sqrt{5}}{2\pi}\right]\) is indicated by dashed black lines.}
	\label{fig:freq_flux_response_flip1}
\end{figure}

\paragraph{Dependence of the work rate on the forcing frequency.}
In Figure~\ref{fig:freq_flux_response_flip1}, we rely on \eqref{eq:finite_size_W} to plot \(\mathbb{W}^n\) as a function of the frequency, for \(T \in \left\{0.1, 1.0, 2.5\right\}\) and forcing magnitude \(f_0 = 1\). Note that the expression in \eqref{eq:finite_size_W} allows for an easy evaluation of the rate of work in a wider range of forcing frequencies. The first particle momentum autocorrelation in the expression of \(\mathbb{W}^n\) is numerically estimated.
As a comparison, we plot in the same figure the rate of work in the harmonic case, which is independent of temperature. At the flip rate values we are considering, the energy flux versus frequency curve of the periodically forced $\beta$-FPUT chain has a shape which is increasingly similar to that of the harmonic chain as the temperature decreases, and they essentially coincide at \(T = 0.1\).

\paragraph{Validity of the PDE describing the temperature profile.}
\begin{figure}[h]
	\centering
	\includegraphics[width = 0.49\textwidth]{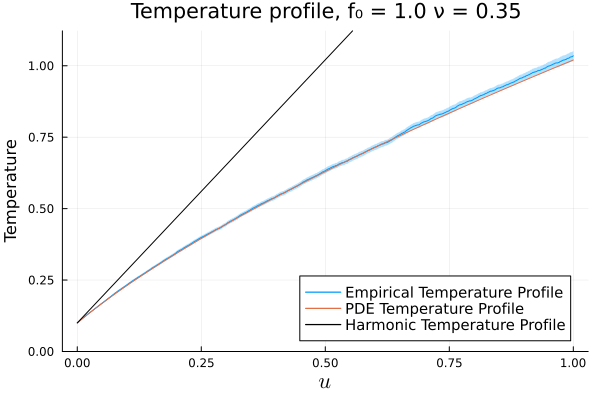}
	\includegraphics[width = 0.49\textwidth]{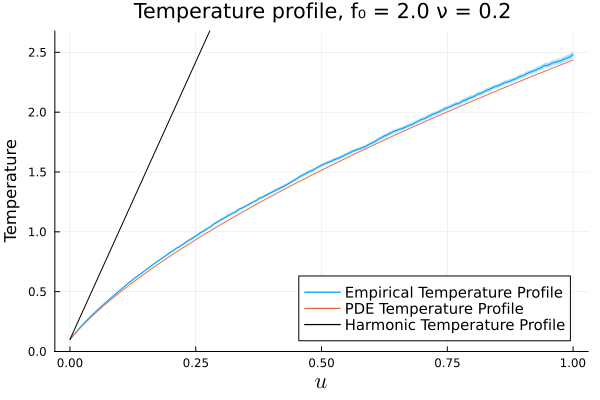}
	\caption{Temperature profile obtained by solving the PDE \eqref{eq:limite_hydro}, compared to the empirical profile computed with the same forcing parameters and flip rate \(\widetilde{\gamma} = 1.0\). Error bars around the empirical profile are shown as a blue shaded region. The harmonic temperature profile for the same forcing parameters is plotted in black.}
	\label{fig:temp_profile_flip1}
\end{figure}
In Figure~\ref{fig:temp_profile_flip1}, we compare the empirically observed temperature profiles obtained by \eqref{eq:temperature_estimator} with the corresponding solutions of the expected limiting PDE \eqref{eq:limite_hydro}. We also plot the linear temperature profile of the harmonic chain given by \cite[Theorem 3.4]{KLO1}. Beyond the low--temperature regime, the anharmonic and harmonic temperature profiles diverge from one another.
With only minor discrepancies, the simulated temperature profiles and energy fluxes closely match those predicted by the conjectured limiting equations. This validates the relevance of the hydrodynamic limit.

\subsection{Small flip rate \(0 < \widetilde{\gamma} < 1\)}\label{sec:small_flip_rate}
In this subsection, we report the results of simulations with small momentum flip rate values, namely \(\widetilde{\gamma} \in \{0.316,\)\( 0.1,\) \(0.0316,\) \(0.01,\) \(0.00316\}\). This choice is motivated by the fact that \(\left(\sqrt{10}\right)^{-1} \approx 0.316\), so that the set of flip rates that we use is roughly a geometric series that decreases by a factor of \(10\) every two steps. As in the previous subsection, we first compare the energy flux \eqref{eq:bulk_flux_estimator} observed in simulations and the numerical estimate of the rate of work given by \eqref{eq:finite_size_W}, at the temperature empirically observed at the forced atom site. Then we study the relationship between the forcing frequency and the rate of work, paying particular attention to what happens as \(\widetilde{\gamma}\) goes to zero. We conclude the subsection by plotting and comparing the temperature profiles observed in simulations with those obtained by solving the PDE \eqref{eq:limite_hydro}.

\paragraph{Validity of the Green--Kubo like formula.}
As in the previous section, we compare the empirical energy flux \(J_n\) with the estimated rate of work \(\mathbb{W}^n\) at the temperature empirically observed at the forced atom site, see Table~\ref{tab:flip0.1} in Appendix~\ref{sec:correlation} for the values of this temperature. For each value of the flip rate we simulated the periodically forced chain with each of the following forcing frequencies and intensities:
\[\left(\nu, f_0\right) \in \left\{0.15, 0.16, 0.175, 0.2, 0.225, 0.25, 0.275, 0.3, 0.325, 0.35\right\} \times \left\{0.5, 1.0, 2.0, 3.0, 4.0, 5.0\right\}.\]
The results are similar to those obtained for \(\widetilde{\gamma} = 1\): we observe a good agreement between $-J_n$ and $\mathbb W^n$ across all the considered values of flip rates and forcing parameters. 
\begin{figure}[h]
	\centering
	\includegraphics[width = 0.49\textwidth]{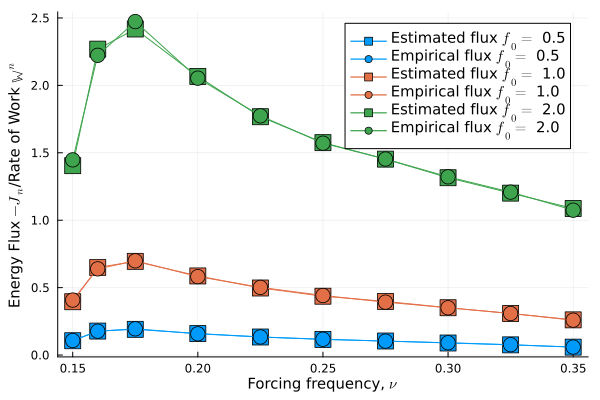}
	\includegraphics[width = 0.49\textwidth]{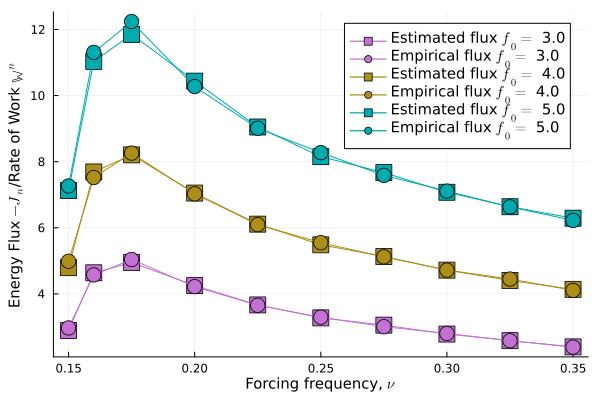}
	\caption{Empirical negative flux observed in simulations with flip rate \(\widetilde{\gamma} = 0.1\) (circles), and work rate estimated using \eqref{eq:finite_size_W} (squares), both plotted against frequency within the harmonic band, using the same forcing parameters and evaluated at the temperature of the forced atom site.}
	\label{fig:emp_flux_v_work_flip01}
\end{figure}
In Figure~\ref{fig:emp_flux_v_work_flip01} we plot the observed energy flux and the estimated rate of work for \(\widetilde{\gamma} = 0.1\) as a representative example.
The empirical energy flux \(-J_n\) closely matches the estimated values of the rate of work \(\mathbb{W}^n\) at the chosen frequencies. We observe a small discrepancy between these two quantities, mainly at large values of the flux and forcing magnitudes where one would expect a larger variance in the estimation of both \(\mathbb{W}^n\) and \(J_n\).
The same results were observed for all the considered values of the flip rates and the corresponding figures may be found in Appendix~\ref{sec:additional_plots}. This agreement suggests that, even in the limit \(\widetilde{\gamma} \to 0\), the expression \eqref{eq:finite_size_W} for \(\mathbb{W}^n\) correctly estimates the asymptotic energy flux/work rate.

\paragraph{Dependence of the work rate on the forcing frequency.}
\begin{figure}[h]
	\centering
	\includegraphics[width = 0.49\textwidth]{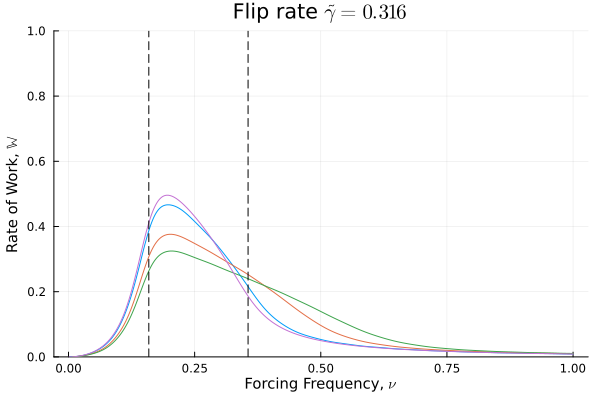}
	\includegraphics[width = 0.49\textwidth]{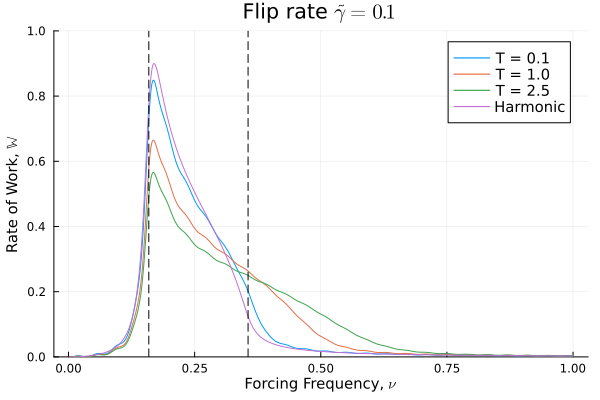}
	\caption{The work rate estimated using \eqref{eq:finite_size_W} as a function of the forcing frequency \(\nu\) with \(f_0 = 1\). The harmonic band of frequencies is marked with dashed black lines.}
	\label{fig:freq_flux_response_smallflip}
\end{figure}
\begin{figure}[h]
	\centering
	\includegraphics[width = 0.49\textwidth]{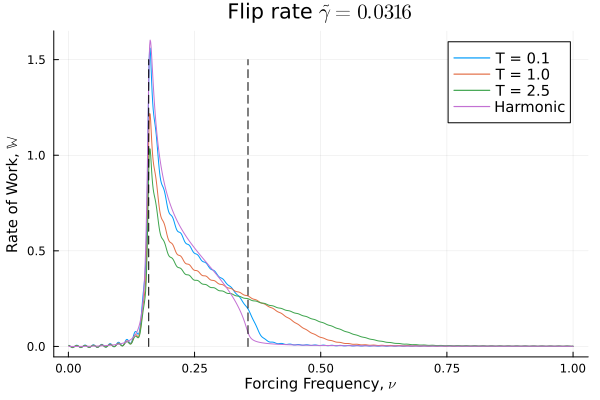}
	\includegraphics[width = 0.49\textwidth]{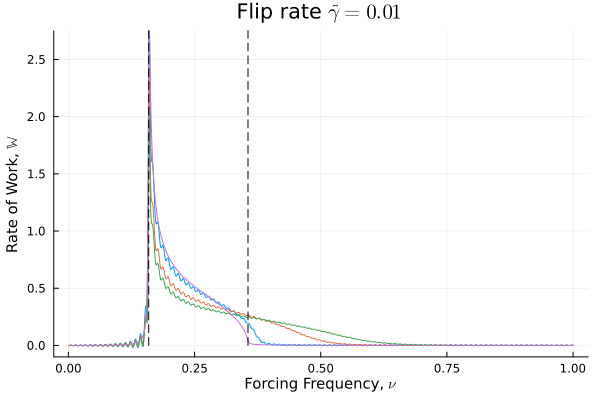}
	\includegraphics[width = 0.49\textwidth]{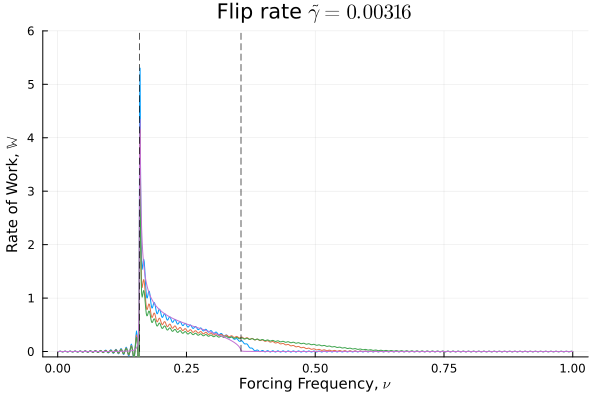}
	\caption{Work rate estimated using \eqref{eq:finite_size_W} as a function of the forcing frequency \(\nu\) with \(f_0 = 1\). The harmonic band of frequencies is marked with dashed black lines.}
	\label{fig:freq_flux_response_vsmallflip}
\end{figure}
In Figures~\ref{fig:freq_flux_response_smallflip} and~\ref{fig:freq_flux_response_vsmallflip}, the rate of work \(\mathbb{W}^n\) is plotted as a function of the forcing frequency for the set of considered values of flip rate, computed using the momentum autocorrelation of the equilibrium chain at temperatures \(T \in \left\{0.1, 1, 2.5\right\}\) and forcing magnitude \(f_0 = 1\). We plot for comparison the rate of work done in the harmonic case. As in the case $\widetilde \gamma = 1$, the frequency response of the anharmonic chain at the lowest temperature \(T_l = 0.1\) is similar to that of the harmonic chain for all the considered values of the flip rate.

Comparing Figures~\ref{fig:freq_flux_response_flip1}, \ref{fig:freq_flux_response_smallflip}~and~\ref{fig:freq_flux_response_vsmallflip}, it appears that, as the flip rate decreases, the response concentrates around the harmonic band \(\left[\widebar{\nu}, \frac{\sqrt{5}}{2\pi}\right]\), which is delimited by dashed black lines in the plots. Below the harmonic band, the response of harmonic and anharmonic chains at the three temperatures behaves similarly: the four curves decay as the forcing frequency goes to zero. This decay appears to be faster at lower flip rate. In all cases considered, the response peaks within the harmonic band, near its lower bound \(\widebar{\nu}\), with more pronounced peaking at lower flip rates. The peaks for the anharmonic chain are lower than those for the harmonic chain, and this difference becomes more evident at higher temperatures.
Above the harmonic band, a more pronounced divergence emerges between the behaviors of the harmonic and anharmonic chains. For these forcing frequencies, both chains exhibit a rapid decay in response at temperature \(T = 0.1\), particularly at lower flip rates. In contrast, anharmonic chains at higher temperatures continue to exhibit a non-negligible response across a range of frequencies above the upper limit of the harmonic band. This frequency range appears to widen as the temperature increases. This phenomenon is known as {\em supratransmission}, and has been reported in the physics literature for various models of periodically forced atomic chains (see \cite{Geniet_Leon, KLR, PBS}). Moreover, the peak at \(\widebar{\nu}\) becomes higher and more pronounced as \(\widetilde{\gamma}\) decreases, suggesting that the rate of work may diverge at \(\widebar{\nu}\) as \(\widetilde{\gamma} \to 0\).

\paragraph{Validity of the PDE describing the temperature profile.}
\begin{figure}[h]
	\centering
	\includegraphics[width = 0.49\textwidth]{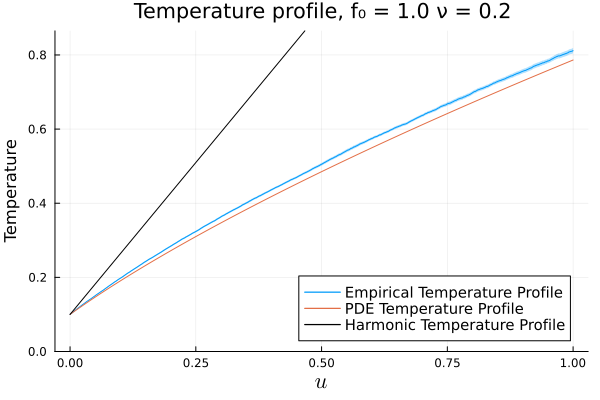}
	\includegraphics[width = 0.49\textwidth]{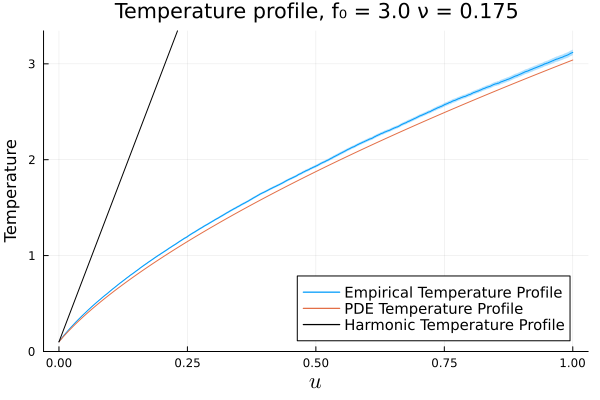}
	\caption{Temperature profile obtained by solving the PDE \eqref{eq:limite_hydro}, compared to the empirical profile computed under the same forcing parameters and flip rate \(\widetilde{\gamma} = 0.316\). Error bars around the empirical profile are shown as a blue shaded region. The harmonic temperature profile for the same forcing parameters is plotted in black.}
	\label{fig:temp_profile_flip0316}
\end{figure}
We conclude this subsection by comparing the empirical temperature profile with the numerical solution of \eqref{eq:limite_hydro}. 
Figure~\ref{fig:temp_profile_flip0316} shows the temperature profiles for flip rate \(\widetilde{\gamma} = 0.316\) and a selection of forcing parameters, presented as a representative example. Temperature profiles for other flip rates exhibit qualitatively similar behavior and are provided in Appendix~\ref{sec:additional_plots} for completeness. As previously noted in the case \(\widetilde{\gamma} = 1\), the empirical temperature profile closely follows the numerical solution of \eqref{eq:limite_hydro}, with some discrepancies. Furthermore, we see a large difference between the linear harmonic temperature profiles and the non--linear anharmonic temperature profiles. In all, we observe good agreement between the empirical temperature profiles and the temperature profiles predicted by \eqref{eq:limite_hydro} as the flip rate decreases. These simulation results support the validity of the hydrodynamic limit in the small flip rate regime.

\subsection{No flip \(\widetilde{\gamma} = 0\)}\label{sec:no_flip}
In this final subsection, we present the results of simulations performed without momentum flip. The mathematical justification for the hydrodynamic limit \eqref{eq:limite_hydro} announced in Section~\ref{sec:theory} is less complete in the case without flip. Consequently, we focus on numerically exploring the behavior of the periodically forced, pinned anharmonic chain without momentum flip. Our results may be compared with those in~\cite{Garrido}, which concern the harmonic chain under the same conditions.
We begin by comparing the scaled energy flux, defined in \eqref{eq:bulk_flux_estimator}, to the rate of work done by the forcing, estimated via formula \eqref{eq:finite_size_W}. We then examine the relationship between the forcing frequency and the rate of work. Finally, we conclude the section by plotting and comparing the temperature profiles obtained from direct simulation with those obtained by numerically solving the PDE \eqref{eq:limite_hydro}.

\begin{figure}[h]
	\centering
	\includegraphics[width = 0.49\textwidth]{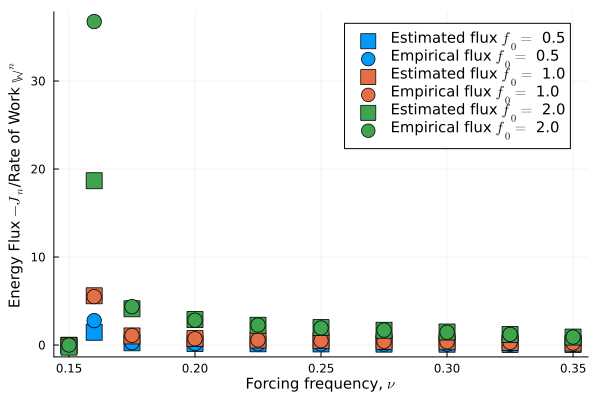}
	\includegraphics[width = 0.49\textwidth]{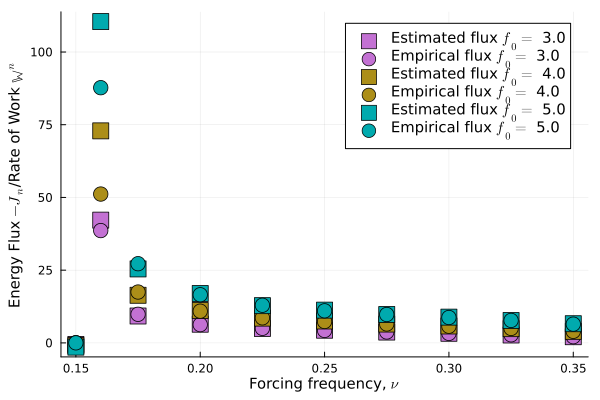}
	\caption{Empirical negative flux observed in simulations without momentum flip (circles), and work rate estimated using \eqref{eq:finite_size_W} (squares), both plotted against frequency within the harmonic band, using the same forcing parameters and evaluated at the temperature of the forced atom site.}
	\label{fig:emp_flux_v_work_noflip}
\end{figure}

\paragraph{Empirical flux versus estimated rate of work.}
The empirically observed flux \(-J_n\) and the estimated rate of work \(\mathbb{W}^n\) are plotted in Figure~\ref{fig:emp_flux_v_work_noflip}.
As before, we simulated the periodically forced chain without momentum flip for each of the following forcing frequencies and intensities 
\[\left(\nu, f_0\right) \in \left\{0.15, 0.16, 0.175, 0.2, 0.225, 0.25, 0.275, 0.3, 0.325, 0.35\right\} \times \left\{0.5, 1.0, 2.0, 3.0, 4.0, 5.0\right\}.\]
In Appendix~\ref{sec:correlation}, Table~\ref{tab:flip0} reports the temperatures observed at the forced atom site and used to estimate the rate of work. The two curves in Figure~\ref{fig:emp_flux_v_work_noflip} exhibit similar overall behaviors: both have a peak at the forcing frequency \(\nu = 0.16\), are nearly zero at \(\nu = 0.15\) --- a frequency below the lower limit of the harmonic band  \(\widebar{\nu}\) --- and decrease as the frequency increases.
Except for the forcing frequency \(\nu = 0.16\), we see good agreement between the rate of work predicted by \eqref{eq:finite_size_W} and the empirically measured flux. At forcing frequency \(\nu = 0.16\), we see quite large discrepancy between the estimated rate of work and the empirically measured flux. In the absence of momentum flip, we observe discontinuities at both ends of the empirical temperature profile. Consequently it is not fully clear at what temperature we should be estimating the rate of work. Nonetheless, the observed discrepancies are much larger than what one would expect from estimating the rate of work at a slightly wrong temperature. See Figure~\ref{fig:work_fit} in Appendix~\ref{sec:correlation} for plots of the rate of work as function of temperature for which it is clear the \(\mathbb{W}\) does not vary dramatically with small temperature differences. This observed difference could be due to the fact that \(\nu\) is close to the lower limit of the harmonic band \(\widebar{\nu}\). The study of the rate of work \(\mathbb{W}\) as function frequency in the previous section for small flip rate and in the next paragraph for no momentum flip suggest that the rate of work has a singularity at \(\widebar{\nu}\) in the hydrodynamic limit. 

\begin{figure}[h]
	\centering
	\includegraphics[width = 0.6\textwidth]{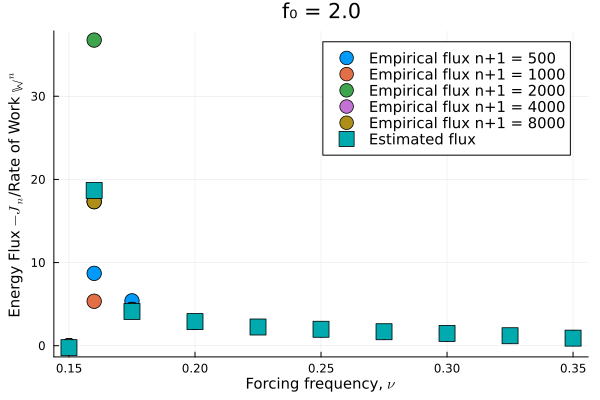}
	\caption{Empirical negative flux observed in simulations without momentum flip (circles) and work rate estimated using \eqref{eq:finite_size_W} (squares), both plotted against frequency within the harmonic band, for the same forcing parameters and evaluated at the temperature of the forced atom.}
	\label{fig:emp_flux_v_work_noflip_by_size}
\end{figure}

To investigate potential finite-size effects, we simulate periodically forced chains of lengths \(500\), \(1000\), \(2000\), \(4000\), \(8000\) at forcing frequencies  \(\nu \in \left\{0.15, 0.16, 0.175, 0.2, 0.225, 0.25, 0.275, 0.3, 0.325, 0.35\right\}\) and forcing intensity \(f_0 = 2.0\). 
The results of these simulations are shown in Figure~\ref{fig:emp_flux_v_work_noflip_by_size}, along with the same estimated  rate of work values for \(f_0 = 2.0\), which are plotted in Figure~\ref{fig:emp_flux_v_work_noflip}.
For forcing frequencies \(\nu = 0.15\) and \(\nu \geq 0.2\), we observe largely consistent values of the scaled energy flux \(J_n\) across different system sizes, and these values closely match the estimated rate of work \(\mathbb{W}^n\). 
In contrast, for \(\nu \in \{0.16, 0.175\}\), we observe size-dependent effects. 
At \(\nu = 0.175\) , the empirically measured flux appears to stabilize for chain lengths of around 1000 atoms or more at the estimated rate of work at the same frequency and at the temperature observed at the forced end of the 2000-atom chain. At \(\nu = 0.16\), the  measured flux does not appear to stabilize until at least chain lengths of 4000 atoms, and it is not monotonic with system size. The empirical flux of the chains of length 4000 and 8000 are close to the estimated rate of work. These observations suggest that large system sizes are required to accurately capture the limiting behavior of the energy flux near the lower bound of the harmonic band.

\paragraph{Dependence of the work rate on the forcing frequency.}
We next study \(\mathbb{W}^n\) as a function of the forcing frequency at temperatures \(T \in \{0.1, 1, 2.5\}\). In this setting, we do not plot the rate of work for the harmonic case, since without momentum flip the rate of work in the hydrodynamic limit is either infinite---if the forcing frequency lies within the harmonic band---or zero otherwise \cite{Garrido}.
\begin{figure}[h]
	\centering
	\includegraphics[width = 0.6\textwidth]{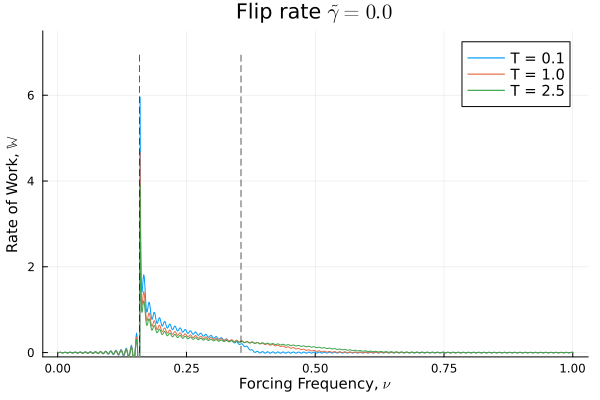}
	\caption{Work rate estimated using \eqref{eq:finite_size_W} plotted against the forcing frequency \(\nu\), with \(f_0 = 1\). The harmonic band 
	\(\nu \in \left[\widebar{\nu}, \frac{\sqrt{5}}{2\pi}\right]\) is indicated by dashed black lines.}
	\label{fig:freq_flux_response_noflip}
\end{figure}

At the upper bound of the harmonic band, we again observe supratransmission. The frequency interval over which a nonzero response is observed increases with temperature. 
As we already remarked in the introduction, this is in contrast with the case with bounded (small) anharmonicity where the absence of supratransmission is rigorously proven in \cite{Garrido2}. 
At the lower bound of the harmonic bound, the functional form of \(\widehat{C}_{p, 4}(t, K, \nu^\star, \phi)\) used to fit the tail of the correlation function leads to a divergence of \(\mathbb{W}^n\) \(\nu\) approaches \(\nu^\star\) from above. When fitting the parameters of \(\widehat{C}_{p, 4}\) in \eqref{eq:Cp4}, we found that \(\nu^\star \approx \widebar{\nu}\) to three decimal places. This indicates that the energy flux through the anharmonic chain should diverge as the forcing frequency \(\widebar{\nu}\), due to a resonance phenomenon. The apparent singularity at \(\widebar{\nu}\) in the system’s response to periodic forcing suggests treating separately the two cases \(\nu \approx \widebar{\nu}\) and \(\nu \gg \widebar{\nu}\).
\paragraph{Validity of the PDE describing the temperature profile.}

The analysis of the rate of work as a function of the forcing frequency suggests that two distinct regimes should be considered separately: the near-resonant case \(\nu \approx \widebar{\nu}\) and the high-frequency regime \(\nu \gg \widebar{\nu}\).

\begin{figure}[h]
	\centering
	\includegraphics[width = 0.6\textwidth]{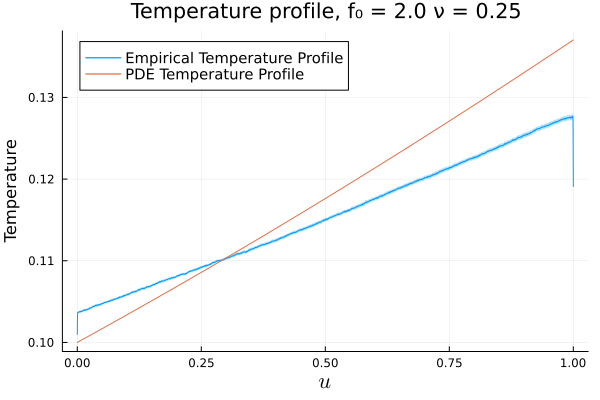}
	\caption{Temperature profile obtained by solving the PDE \eqref{eq:limite_hydro}, compared to the empirical profile computed under the same forcing parameters and a forcing frequency away from \(\widebar{\nu}\), in the absence of momentum flip. Error bars around the empirical profile are shown as a blue shaded region.}
	\label{fig:temp_profile_flip0}
\end{figure}

We begin by examining forcing frequencies \(\nu \geq 0.2\). In this range, the stationary temperature profiles obtained from simulations are approximately linear and increase from left to right. However, they differ significantly from the profiles produced by numerically solving the PDE \eqref{eq:limite_hydro}; see Figure~\ref{fig:temp_profile_flip0} for a representative example. We do not plot the harmonic temperature profile in Figure~\ref{fig:temp_profile_flip0} since it is flat when there is no momentum flip \cite{Garrido}. Although both profiles are nearly linear, their slopes differ markedly. Furthermore, the empirical temperature profiles display boundary jumps (due to thermal boundary resistances\footnotemark\footnotetext{The emergence of a boundary temperature jump reflects a resistance to the transfer of thermal energy across the interface between the system and the thermostat. This resistance arises from inefficient phonon transmission, due to reflection or scattering at the microscopic level.}, see \eg~\cite{AK01}), which are not compatible with solutions of \eqref{eq:limite_hydro}, as the latter are continuously differentiable \(C^1\).

\begin{figure}[h]
	\centering
	\includegraphics[width = 0.6\textwidth]{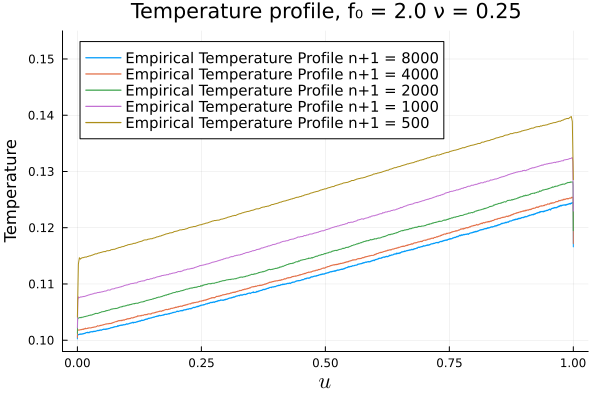}
	\caption{The empirically observed temperature profile for atom chains of different lengths with no momentum flip and forcing frequency away from \(\widebar{\nu}\).}
	\label{fig:temp_profile_flip0_by_size}
\end{figure}

To assess the impact of system size, we plot in Figure~\ref{fig:temp_profile_flip0_by_size} the empirical temperature profiles for chains of lengths 500, 1000, 2000, 4000, and 8000 under the same forcing parameters as in Figure~\ref{fig:temp_profile_flip0}. Across all sizes, the bulk of the profile exhibits the same slope, while the differences between the curves arise primarily from boundary discontinuities. These jumps diminish as the system size increases. Such behavior has been reported in other atom chain models lacking bulk noise; see, for example, \cite{LLP}. In particular, the temperature at the left boundary tends to approach the thermostat temperature \(T_\ell = 0.1\) as the system size grows.

\begin{figure}[h]
	\centering
	\includegraphics[width = 0.49\textwidth]{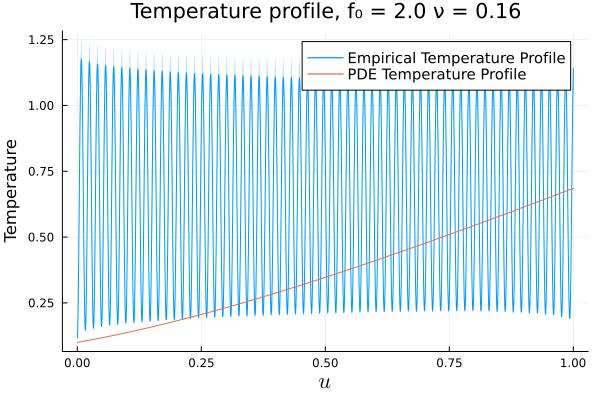}
	\includegraphics[width = 0.49\textwidth]{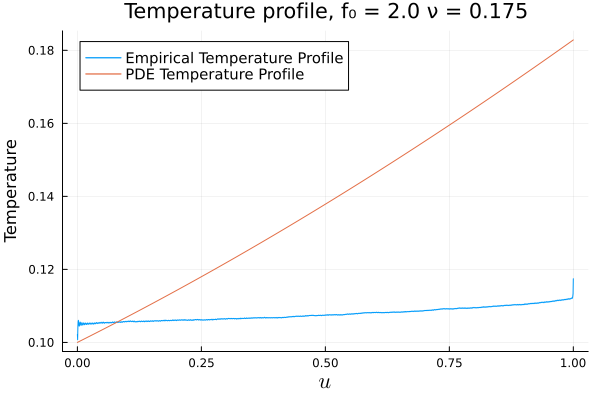}
	\caption{Same as Figure~\ref{fig:temp_profile_flip0} for forcing frequencies close to \(\widebar{\nu}\).}
	\label{fig:temp_profile_flip0_nearsingularity}
\end{figure}

We now turn to forcing frequencies close to the apparent singularity, specifically \(\nu \in \{0.16, 0.175\}\); see Figure~\ref{fig:temp_profile_flip0_nearsingularity}. In this regime, we observe strong deviations between the numerical solutions of \eqref{eq:limite_hydro} and the empirical temperature profiles. For \(\nu = 0.16\), very close to \(\widebar{\nu} \approx 0.159\), the simulated temperature profile is spatially oscillatory and periodic. Such behavior cannot correspond to a mild solution of the PDE \eqref{eq:limite_hydro}, as any such solution must have a non-negative derivative; see \eqref{eq:fixed_point_problem}. For \(\nu = 0.175\), the empirical profile is nearly flat, with only small jumps at the boundaries -- again in stark contrast to the increasing profiles predicted by the PDE.

\begin{figure}[h]
	\centering
	\includegraphics[width = 0.49\textwidth]{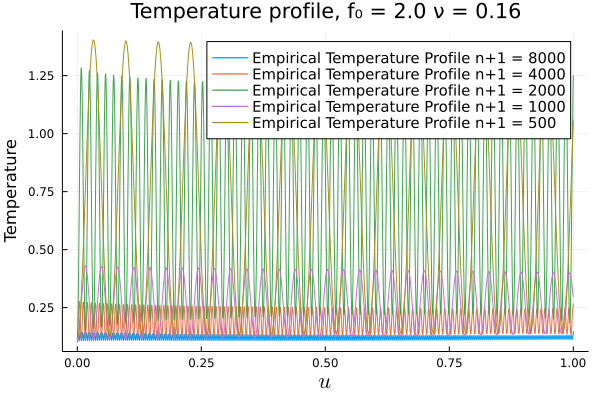}
	\includegraphics[width = 0.49\textwidth]{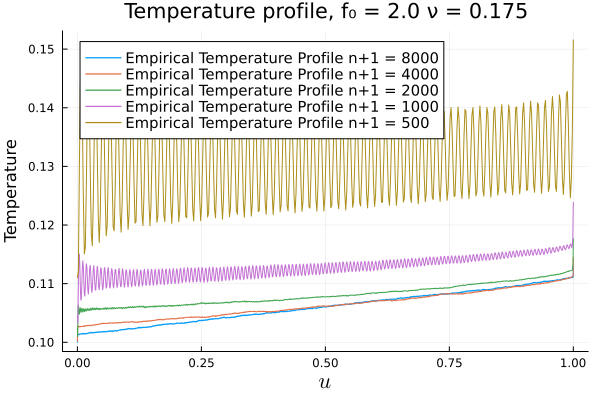}
	\caption{Same as Figure~\ref{fig:temp_profile_flip0_by_size} for forcing frequencies close to \(\widebar{\nu}\).}
	\label{fig:temp_profile_flip0_nearsingularity_by_size}
\end{figure}

To better understand finite--size effects in this near--resonant regime, we plot in Figure~\ref{fig:temp_profile_flip0_nearsingularity_by_size} the empirical temperature profiles for chains of lengths 500, 1000, 2000, 4000, and 8000, again using the same forcing parameters as in Figure~\ref{fig:temp_profile_flip0_nearsingularity}. At \(\nu = 0.16\), all profiles exhibit spatially periodic oscillations, whose wavelength appears inversely proportional to the system size: doubling the chain length roughly halves the spatial oscillation period.

The oscillation amplitude does not vary monotonically with system size. While the smallest system (length 500) shows the largest amplitude, the 2000-atom chain displays larger oscillations than the 1000-atom one. For \(\nu = 0.175\), oscillations are present in the two smallest systems, but their amplitude decreases with system size. For chains of length 2000 and above, the profiles become approximately linear with small slope and decreasing boundary jumps.

\section{Conclusion}\label{sec:conclusions}
Our main objective in this work was to numerically investigate whether and how the $\beta$-FPUT chain with harmonic pinning and momentum-flip noise exhibits the same type of hydrodynamic, diffusive-scale behavior as its harmonic counterpart under time-periodic boundary forcing. In particular, we focused on verifying that, in the long-chain, large-time limit (the hydrodynamic limit), the system’s temperature profile converges to the solution of a heat-type partial differential equation, with a Dirichlet boundary condition on the thermostatted end and a Neumann-type flux condition on the periodically forced end.

We also examined whether the average energy current in the steady state could be described through a Green–Kubo-type formula, as rigorously established for harmonic chains in~\cite{KLO1,KLO2}. Although no closed-form expression is available in the anharmonic case, our numerical results support this formulation: both the thermal conductivity and the work-induced flux can be consistently estimated from simulations, and used as inputs in the macroscopic PDE, leading to excellent agreement with the observed temperature profiles.

Beyond confirming the emergence of diffusive behavior at the macroscopic scale, our results reveal nonlinear energy transport features absent in the harmonic case. In particular, we observe supratransmission phenomena for forcing frequencies above the harmonic band, even in the absence of bulk stochasticity. We also identify resonance-like behavior near the lower edge of the harmonic spectrum as the flip rate decreases, and show that the transmission band depends on temperature, highlighting the role of anharmonicity in enabling these effects.

A natural direction for further investigation is the extension of the present study to systems with multiple conserved quantities, such as energy, momentum, or volume stretch. Recent rigorous results~\cite{KOS24} on the hydrodynamic limit of the unpinned, periodically driven harmonic chain in the presence of velocity-flip noise have revealed the interplay between mechanical and thermal energy flows, as well as the emergence, in the diffusive limit, of a system of coupled macroscopic PDEs for the conserved fields, subject to mixed Dirichlet- and Neumann-type boundary conditions. Whether these macroscopic features persist in nonlinear settings remains an open question from a mathematical standpoint. In this context, it would be particularly relevant to numerically investigate the unpinned case of the model studied here, or a periodically driven rotor chain (with conservation of energy and momentum), in order to test the conjecture that the qualitative and quantitative macroscopic behavior established in the harmonic case with flip dynamics remains valid in the presence of anharmonic interactions. Even a numerical confirmation of this conjecture would provide valuable insight into the robustness of diffusive transport and boundary-induced effects in nonlinear, low-dimensional systems.

\appendix

\section{Derivation of the expression \eqref{eq:gk_work} of $\mathbb W$}\label{sec:gk_work_deriv}
In this appendix section, we provide a heuristic derivation of \eqref{eq:gk_work} using a linear-response type argument and comment on how it could be rigorously obtained. 
The rate of work done by the forcing on the last particle at temperature \(T\) is given by
\[\mathbb{W}\left(T, \mathdutchcal{F}, \theta\right) = -\, \lim_{n \to \infty} n \llangle j_{n, n+1}\rrangle_{n}.\]
We suppose that \(\rho_n\) admits a density \(\psi_n\left(t, q, p\right)\) with respect to \(\nu_{T, \theta}^n\left(\dd t\, \dd q\, \dd p\right) := \frac{1}{\theta}\,\dd t \otimes \mu_T^n\left(\dd q\, \dd p\right)\). The density \(\psi_n\) is the \(\theta\)-periodic solution of the Fokker--Planck equation
\[\left(-\partial_t + \mathdutchcal{L}_0^* + \frac{1}{\sqrt{n}}\widetilde{\mathdutchcal{L}}_t^*\right)\psi_n = 0,\]
where adjoints are taken in \(L^2\left(\nu_{T, \theta}^n\right)\), the space \(\theta\)--time--periodic functions that are square integrable with respect to \(\nu_{T, \theta}^n\) over a \(\theta\)--period. This suggests to look for \(\psi_n\) as a series expansion in \(\frac{1}{\sqrt{n}}\), namely
\begin{equation}\label{eq:psi_expansion}
	\psi_n(t,q,p) = 1 + \frac{1}{\sqrt{n}}\mathfrak{f}_n^1(t,q,p) + \mathrm{O}\left(\frac{1}{n}\right).
\end{equation}
By matching first order terms, one finds that \(\mathfrak{f}_n^1\) solves
\[-\left(-\partial_t + \mathdutchcal{L}_0^*\right)\mathfrak{f}_n^1 = \widetilde{\mathdutchcal{L}}_t^* \mathbf{1} = \mathdutchcal{F}\left(\frac{t}{\theta}\right)\frac{p_n}{T}.\]
We denote by \(\mathbb{E}_T^n\) the expectation with respect to the equilibrium dynamics of a finite chain with free boundary conditions:
\begin{equation}\label{eq:finite_eq_chain}
	\begin{aligned}
		d q_x &= p_x \, \dd t, && x \in \left\{0, 1, \dots, n\right\},\\
		\dd p_0 &= \left(v'\left(q_1 - q_0\right) - u'\left(q_0\right)\right)\dd t -2p_0\dd N_0\left(\widetilde{\gamma}t\right),\\
		\dd p_x &= \left(v'\left(q_{x+1} - q_x\right) - v'\left(q_x - q_{x-1}\right) - u'\left(q_x\right)\right)\dd t - 2p_x\dd N_x\left(\widetilde{\gamma}t\right) && x \in \left\{1, \dots, n-1\right\},\\
		\dd p_n &= \left(-v'\left(q_n - q_{n-1}\right) - u'\left(q_n\right)\right)\dd t -2p_n\dd N_n\left(\widetilde{\gamma}t\right),\\
	\end{aligned}
\end{equation}
started at the Boltzmann--Gibbs measure \(\mu_T^n\) at temperature \(T\). In view of \eqref{eq:bd_flux_def}, we compute
\begin{align*}
		-n \llangle j_{n, n+1}\rrangle_{n} &=  n \frac{1}{\theta}\int_{0}^\theta \int_{\mathbb{R}^{n+1} \times \mathbb{R}^{n+1}} \frac{1}{\sqrt{n}} \mathdutchcal{F}\left(\frac{s}{\theta}\right)p_n \psi_n\left(s, q, p\right) \mu_T^n\left(\dd q\,\dd p\right)\dd s.\\
		\intertext{Using the series expansion \eqref{eq:psi_expansion} for \(\psi_n\), the right hand side becomes}
		-n \llangle j_{n, n+1}\rrangle_{n} &= n \frac{1}{\theta}\int_{0}^\theta \int_{\mathbb{R}^{n+1} \times \mathbb{R}^{n+1}} \frac{1}{\sqrt{n}} \mathdutchcal{F}\left(\frac{s}{\theta}\right)p_n\left(1 + \frac{1}{\sqrt{n}}\mathfrak{f}_n^1\left(s, q, p\right) + \mathrm{O}\left(\frac{1}{n}\right)\right)\mu_T^n\left(\dd q\,\dd p\right)\dd s \\
		&= \frac{1}{\theta}\int_{0}^\theta \int_{\mathbb{R}^{n+1} \times \mathbb{R}^{n+1}}  \mathdutchcal{F}\left(\frac{s}{\theta}\right)p_n \mathfrak{f}_n^1\left(s, q, p\right) \mu_T^n\left(\dd q\,\dd p\right)\dd s + \mathrm{O}\left(\frac{1}{\sqrt{n}}\right)\\
		&= \frac{1}{\theta}\int_{0}^\theta \int_{\mathbb{R}^{n+1} \times \mathbb{R}^{n+1}}  \mathdutchcal{F}\left(\frac{s}{\theta}\right)p_n \left[-\left(-\partial_s + \mathdutchcal{L}_0^*\right)^{-1}\left(\mathdutchcal{F}\left(\frac{s}{\theta}\right)\frac{p_n}{T}\right)\right] \mu_T^n\left(\dd q\,\dd p\right)\dd s + \mathrm{O}\left(\frac{1}{\sqrt{n}}\right).
		\intertext{The adjoint in \(L^2\left(\nu_{T, \theta}^n\right)\) of \(-\partial_s + \mathdutchcal{L}_0^*\) is \(\partial_s + \mathdutchcal{L}_0\), so}
		-n \llangle j_{n, n+1}\rrangle_{n} &= \frac{1}{\theta}\int_{0}^\theta \int_{\mathbb{R}^{n+1} \times \mathbb{R}^{n+1}}  \left[-\left(\partial_s + \mathdutchcal{L}_0\right)^{-1}\mathdutchcal{F}\left(\frac{s}{\theta}\right)p_n\right] \left(\mathdutchcal{F}\left(\frac{s}{\theta}\right)\frac{p_n}{T}\right) \mu_T^n\left(\dd q\,\dd p\right)\dd s + \mathrm{O}\left(\frac{1}{\sqrt{n}}\right).\\
		\intertext{When acting on a function of the form \(f(s,q,p) = h(s)g(q,p)\) where \(h\) depends only on \(s\) and \(g\) depends only on \(q\) and \(p\) and has mean zero with respect to \(\mu_T^n\), the inverse of \(-\left(\partial_s + \mathdutchcal{L}_0\right)\) can be formally written as:
		\[-\left(\partial_s + \mathdutchcal{L}_0\right)^{-1}h(s)g(q,p) = \int_0^\infty h\left(s+t\right) \left(\mathrm{e}^{t\mathdutchcal{L}_0}g\right)(q,p) \dd t,\]
		thus}
		-n \llangle j_{n, n+1}\rrangle_{n} &= \frac{1}{\theta}\int_{0}^\theta \int_{\mathbb{R}^{n+1} \times \mathbb{R}^{n+1}}  \left[\int_0^\infty\mathdutchcal{F}\left(\frac{s+t}{\theta}\right)\mathrm{e}^{t\mathdutchcal{L}_0}p_n\dd t\right] \mathdutchcal{F}\left(\frac{s}{\theta}\right)\frac{p_n}{T} \mu_T^n\left(\dd q\,\dd p\right)\dd s + \mathrm{O}\left(\frac{1}{\sqrt{n}}\right)\\
		&= \frac{1}{T}\int_0^\infty \left(\frac{1}{\theta}\int_0^\theta \mathdutchcal{F}\left(\frac{s + t}{\theta}\right) \mathdutchcal{F}\left(\frac{s}{\theta}\right)\dd s\right) \int_{\mathbb{R}^{n+1} \times \mathbb{R}^{n+1}} p_n\left(\mathrm{e}^{t\mathdutchcal{L}_0}p_n\right) \mu_T^n\left(\dd q\, \dd p\right)\dd t + \mathrm{O}\left(\frac{1}{\sqrt{n}}\right)\\
		&= \frac{1}{T}\int_0^\infty \left(\frac{1}{\theta}\int_0^\theta \mathdutchcal{F}\left(\frac{s + t}{\theta}\right) \mathdutchcal{F}\left(\frac{s}{\theta}\right)\dd s\right) \mathbb{E}_T^n\left[p_n(t)p_n(0)\right]\dd t + \mathrm{O}\left(\frac{1}{\sqrt{n}}\right).\\
		\intertext{By the symmetry of \eqref{eq:finite_eq_chain}, we can replace \(\mathbb{E}_T^n\left[p_n(t)p_n(0)\right]\) with \(\mathbb{E}_T^n\left[p_0(t)p_0(0)\right]\) to arrive at}
		-n \llangle j_{n, n+1}\rrangle_{n} &= \frac{1}{T}\int_0^\infty \left(\frac{1}{\theta}\int_0^\theta \mathdutchcal{F}\left(\frac{s + t}{\theta}\right) \mathdutchcal{F}\left(\frac{s}{\theta}\right)\dd s\right) \mathbb{E}_T^n\left[p_0(t)p_0(0)\right]\dd t + \mathrm{O}\left(\frac{1}{\sqrt{n}}\right).
\end{align*}
Following \cite[Section 4]{Cedric_Stefano} or \cite[Section 2.2]{AlessandraStefanoGabriel}, we replace \(\mathbb{E}_T^n\) with \(\mathbb{E}_T^+\) to take the thermodynamic limit \(n \to \infty\). Heuristically, the dynamics in \eqref{eq:finite_eq_chain} approaches that in \eqref{eq:semi_infinite_eq_chain} as \(n\) gets larger. Consequently, taking the limit \(n \to \infty\) gives
\[\mathbb{W}\left(T, \mathdutchcal{F}, \theta\right) = \frac{1}{T}\int_0^\infty \left(\frac{1}{\theta}\int_0^\theta \mathdutchcal{F}\left(\frac{s + t}{\theta}\right)\mathdutchcal{F}\left(\frac{s}{\theta}\right)\dd s \right) \mathbb{E}_T^+\left[p_0(t)p_0(0)\right]\dd t,\]
which is the desired equality. 

\section{Computation of Correlation Functions and Thermal Conductivity}\label{sec:correlation}
We detail, in this appendix, the numerical estimate of the equilibrium time-correlation functions of the first particle momentum, \(p_0\), and of the sum of the bulk fluxes,
\[J_n = \sum_{x=0}^{n-1}j_{x, x+1},\]
which we use to compute the rate of work done by the forcing \eqref{eq:finite_size_W} and the thermal conductivity~\eqref{eq:thermalcond}. The computation of the work rate was outlined in Section~\ref{sec:numerics} and we make the computation of the conductivity precise at the end of this appendix. As stated in Section~\ref{sec:numerics}, the correlation functions are estimated using simulations of the finite-length chain \eqref{eq:finite_eq_chain}. We simulate \(R\) realizations of the dynamics of length \(L\) time steps by cutting a long trajectory into \(R\) segments of length \(L\).  We denote by \(p_0^{r, \ell}\) and \(J^{r, \ell}\) the instantaneous momentum of the first particle and the sum of the bulk fluxes at the \(\ell\)-th time step of the \(r\)-th realization. The numerical estimators for the momentum and flux correlation functions, denoted respectively by \(\widehat{\mathtt{C}}_p\) and \(\widehat{\mathtt{C}}_J\), are obtained by the usual formulae based on time averages:
\begin{equation}\label{eq:num_p_corr}
	\widehat{\mathtt{C}}_p[\ell] = \frac{1}{R\left(L - \ell\right)} \sum_{r = 1}^{R}\sum_{k = 0}^{L -1 -\ell}p_0^{r, k}p_0^{r, k+\ell} - \left(\frac{1}{RL}\sum_{r=1}^R \sum_{k = 0}^{L-1}p_0^{r,k}\right)^2, \qquad \ell \in \left\{0, \dots, L - 1\right\},
\end{equation}
and
\begin{equation}\label{eq:num_J_corr}
	\widehat{\mathtt{C}}_J[\ell] = \frac{1}{R\left(L - \ell\right)} \sum_{r = 1}^R\sum_{k = 0}^{L - 1 - \ell} J^{r, k}J^{r, k + \ell} - \left(\frac{1}{RL}\sum_{r = 1}^R \sum_{k=0}^{L-1}J^{r,k}\right)^2, \qquad \left\{0, \dots, L - 1\right\}.
\end{equation}
After computing \(\widehat{\mathtt{C}}_p\) and \(\widehat{\mathtt{C}}_J\), we fit functions to their tails to estimate the longtime behavior of the momentum and flux correlation functions. The functional forms~\eqref{eq:Cp1}-\eqref{eq:Cp4} used for fitting to the tail of momentum correlation are discussed in Section~\ref{sec:numerics}. As for the flux, we fit an exponential tail to the correlation function 
\begin{equation}\label{eq:CJ}
    \widehat{C}_J (t; K, \lambda) = K \mathrm{e}^{-\lambda t}.
\end{equation}

\begin{figure}[h]
	\centering
	\includegraphics[width = 0.49\textwidth]{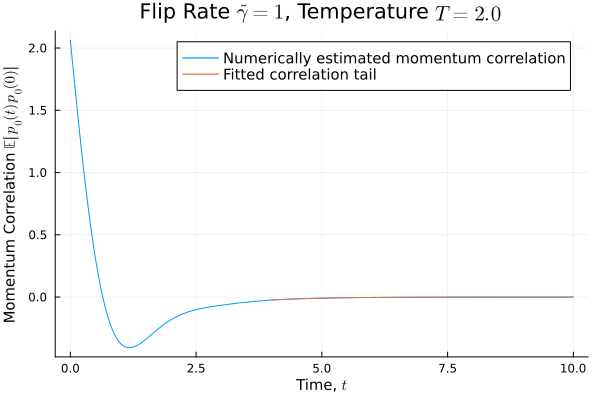}
	\includegraphics[width = 0.49\textwidth]{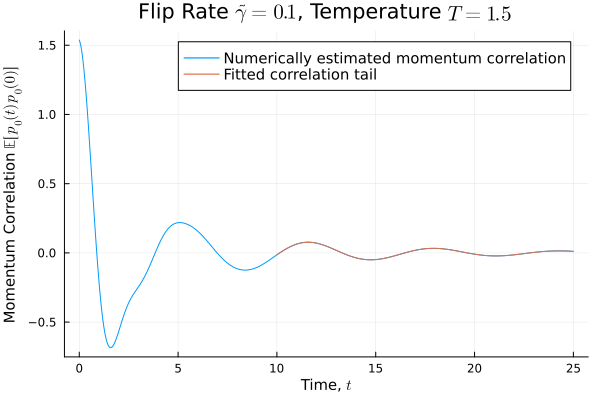}
	\includegraphics[width = 0.49\textwidth]{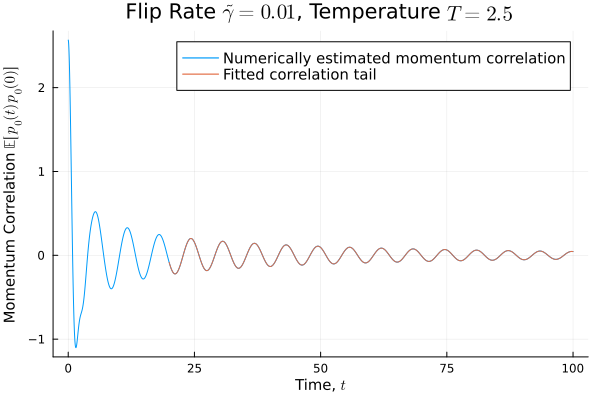}
	\includegraphics[width = 0.49\textwidth]{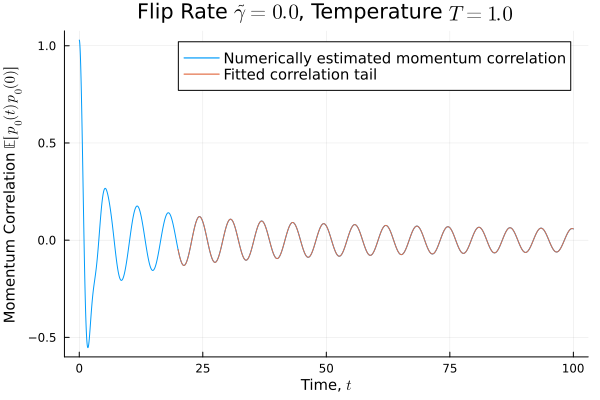}
    \caption{Simulated momentum correlation functions (blue) alongside their fitted curves (orange).}
	\label{fig:momentum_corr_fit}
\end{figure}

We approximate the conductivity by directly adapting the formula \eqref{eq:thermalcond} to the case of finite chain and replacing the integral with a numerical approximation. Thus, to compute the conductivity, we directly integrate the bulk flux correlation function with the trapezoidal rule up to some truncation time \(t_0 = \Delta t \ell_0\) and then analytically integrate the tail based on the function that was fitted:
\begin{equation}\label{eq:Dn}
    D^n(T) = \frac{\Delta t}{2T^2 (n+1)}\sum_{k = 1}^{\ell_0}\left(\widehat{\mathtt{C}}_J[k] + \widehat{\mathtt{C}}_J[k - 1]\right) + \frac{\widehat{C}_J\left(\ell_0 \Delta t; K, \lambda\right)}{\lambda T^2 (n+1)}.
\end{equation}
We fit the parameters of \(\widehat{C}_J\) by minimizing the loss function
\begin{equation}\label{eq:Loss}
\mathrm{Loss}\left(\widehat{\mathtt{C}_J}, \widehat{C}_J; \ell_{\mathrm{start}}, \ell_{\mathrm{end}} \right) = \sum_{k = \ell_{\mathrm{start}}}^{\ell_{\mathrm{end}}} \left(\widehat{\mathtt{C}_J}[k] - \widehat{C}_J\left(\Delta t k\right)\right)^2,
\end{equation}
using a gradient descent in the parameters. The range of time steps over which this loss is computed \(\left\{\ell_{\mathrm{start}}, \ell_{\mathrm{start}} + 1, \dots \ell_{\mathrm{end}}\right\}\) is a parameter to be set. The start of this interval \(\ell_{\mathrm{start}}\) must be chosen sufficiently large to accurately capture the tail behavior, but not too large, to avoid values of \(\widehat{\mathtt{C}}[\ell_{\mathrm{start}}]\) that are too small and dominated by noise. Figure~\ref{fig:momentum_corr_fit} displays some momentum correlation functions along with their respective tail fits.

The autocorrelation functions is estimated using the following parameters: the number of realizations \(R\), the number of time steps per realization \(L\), and the length of the simulated chain \(n + 1\). We chose \(R = 10^5\) realizations, each of duration \((L- 1)\Delta t = 1000\), with a fixed time step \(\Delta t = 0.005\). We observed that this duration was sufficient for capturing short-time correlation behavior and fitting tails to characterize long-time correlations. Preliminary simulations showed minimal differences between chain lengths \(n=1000\) and \(n=2000\), so we selected \(n=1000\) for the final simulations.

To determine the interval for fitting the tail of the correlation functions as well as the truncation times, we performed preliminary simulations to select suitable values for \(\ell_{\mathrm{start}}\), \(\ell_{\mathrm{end}}\), and \(\ell_0\). Our results are reported in Tables~\ref{tab:momentum_fit_intervals} and~\ref{tab:flux_fit_intervals}.

\begin{table}[ht]
\centering
\begin{tabular}{|lcrrcrr|}
\hline
\rowcolor{gray!10}&  &&&&&\\[-0.5em]
\rowcolor{gray!10} $\widetilde{\gamma}$ & fitting function &\(t_0\) &\(\ell_0\) &$[t_\mathrm{start},\,t_\mathrm{end}]$ &$\ell_{\mathrm{start}}$ & $\ell_{\mathrm{end}}$ \\[0.5em]
\hline\hline
&  &&&&&\\[-0.2em]
1.0     &   $\widehat C_{p,1}$  & 8   & 1600  &$[4,\,8]$     &800   & 1600   \\[0.5em]
0.316   &   $\widehat C_{p,2}$  & 15  & 3000  &$[5,\,15]$    &1000  & 3000   \\[0.5em]
0.1     &   $\widehat C_{p,2}$  & 25  & 5000  &$[10,\,25]$   &2000  & 5000   \\[0.5em]
0.0316  &   $\widehat C_{p,3}$  & 60  & 12000 &$[20,\,60]$   &4000  & 12000  \\[0.5em]
0.01    &   $\widehat C_{p,3}$  & 100 & 20000 &$[20,\,100]$  &4000  & 20000  \\[0.5em]
0.00316 &   $\widehat C_{p,3}$  & 100 & 20000 &$[20,\,100]$  &4000  & 20000  \\[0.5em]
0.001   &   $\widehat C_{p,3}$  & 100 & 20000 &$[20,\,100]$  &4000  & 20000  \\[0.5em]
0.0     &   $\widehat C_{p,4}$  & 100 & 20000 &$[20,\,100]$  &4000  & 20000  \\[0.5em]
\hline
\end{tabular}
\caption{Fitting intervals for the tail of the momentum correlation functions for the various values of~$\widetilde\gamma$. The fitting functions are defined in~\eqref{eq:Cp1}-\eqref{eq:Cp4}. }
\label{tab:momentum_fit_intervals}
\end{table}

\begin{table}[ht]
\centering
\begin{tabular}{|lcrrcrr|}
\hline
\rowcolor{gray!10}& & & & & &\\[-0.5em]
\rowcolor{gray!10} $\widetilde{\gamma}$ &  $T$ & $t_{0}$ & $\ell_{0}$ & $[t_\mathrm{start},\,t_\mathrm{end}]$& $\ell_{\mathrm{start}}$ & $\ell_{\mathrm{end}}$ \\[0.5em]
\hline\hline
& & & & &\\[-0.2em]
1       &  all$^{^{(*)}}$& 5   & 1000  & ---& ---& ---\\[0.5em]
0.316   &  $T \ge 1$        & 10  & 2000  & ---& ---& ---\\
        &        $0.8 \le T < 1$  & 11  & 2200  & ---&        ---& ---\\
        &        $T < 0.8$        & 12  & 2400  & ---&        ---& ---\\[0.5em]
0.1     &  $T \ge 0.5$      & 10  & 2000  & $[10,\,20]$& 2000  & 4000\\
        &        $T < 0.5$        & 20  & 4000  & $[20,\,28]$& 4000  &        5600\\[0.5em]
0.0316  &  $T \ge 0.3$      & 20  & 4000  & $[20,\,30]$& 4000  & 6000\\
        &        $T < 0.3$& 40  & 8000  & $[40,\,50]$& 8000  &        10000\\[0.5em]
0.01    &  $T \ge 1$        & 50  & 10000 & $[50,\,75]$& 10000 & 15000\\
        &        $T < 1$          & 75  & 15000 & $[75,\,100]$& 15000 &        20000\\[0.5em]
0.00316 &  all$^{^{(*)}}$& 75  & 15000 & $[75,\,100]$& 15000 & 20000  \\[0.5em]
0.001   &  all$^{^{(*)}}$& 100 & 20000 & $[100,\,125]$& 20000 & 25000\\[0.5em]
0       &  $T \ge 1$        & 100 & 20000 & $[100,\,200]$& 20000 & 40000\\
        &        $T < 1$          & 200 & 40000 & $[200,\,300]$& 40000 &        60000\\[0.3em]
\hline
\end{tabular}\\[0.2em]
{\footnotesize $\phantom{x}^{^{(*)}}$As specified in the main text, temperatures in all simulations range from $T_\mathrm{min} = 0.1$ to $T_\mathrm{max}=3.5$, with increments of 0.1.}
\caption{Fitting intervals and truncation parameters ($\ell_{0},\,t_{0}$) for the estimate of $\widehat{C}_J$ in the tail of the flux correlation functions~\eqref{eq:Dn}, and the corresponding values of~$\widetilde{\gamma}$, at various temperatures. The cases in which the interval $[t_0,\,t_\mathrm{end}]$ is not specified, are cases in which the flux correlation decays very
quickly, thus no tail fitting was performed.}
\label{tab:flux_fit_intervals}
\end{table}

\begin{figure}[h]
	\centering
	\includegraphics[width = 0.49\textwidth]{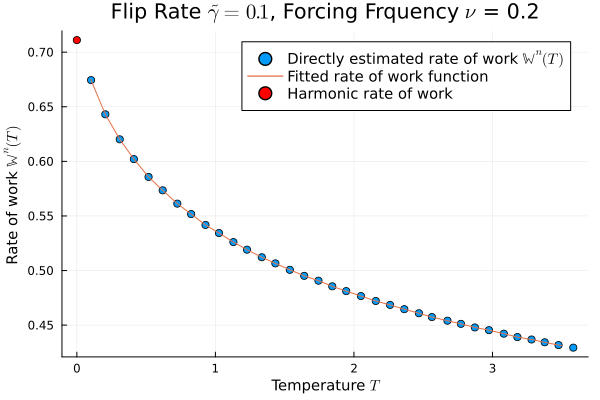}
	\includegraphics[width = 0.49\textwidth]{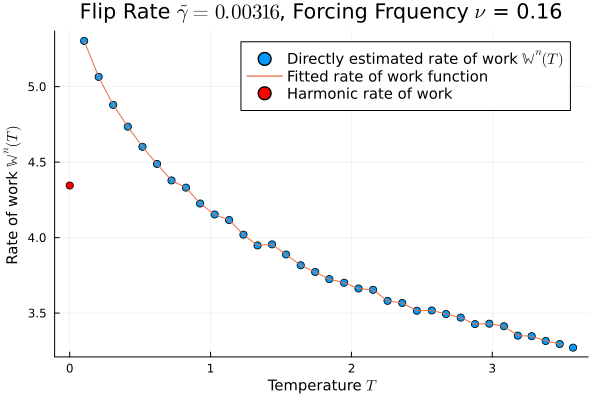}
	\includegraphics[width = 0.49\textwidth]{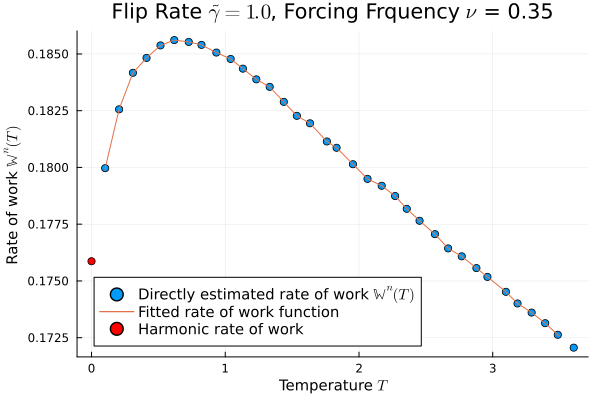}
	\includegraphics[width = 0.49\textwidth]{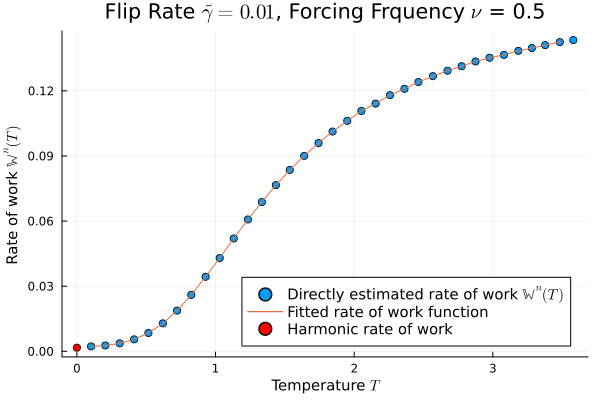}
	\caption{The rate of work computed for the simulated temperatures (blue points) and the function defined by linearly interpolating between these points (orange lines). We plot with a red point at \(T = 0\) the rate of work in the harmonic case, which is temperature independent. All plots are for forcing intensity \(f_0 = 1.0\).}\label{fig:work_fit}
\end{figure}

We repeat the above procedure for temperatures ranging from \( T = 0.1 \) to \( T = 3.5 \) in increments of \( 0.1 \), yielding estimates of \( D^n \) and \( \mathbb{W}^n \) at 35 distinct temperatures. To approximate the function \( T \mapsto D^n(T) \), we fit the data using different functional forms depending on the flip rate. For non-zero flip rates, we include in the fitting procedure the value of the harmonic chain conductivity at \( T = 0 \), computed via \cite[Equation (4.18)]{BLL}. When the flip rate is zero, the conductivity of the harmonic chain is infinite.
\begin{figure}[h]
	\centering
	\includegraphics[width = 0.49\textwidth]{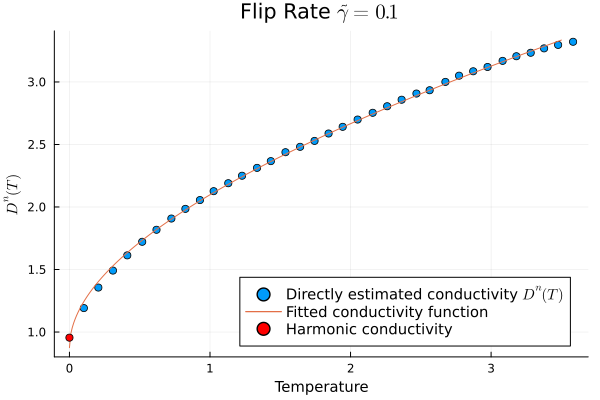}
	\includegraphics[width = 0.49\textwidth]{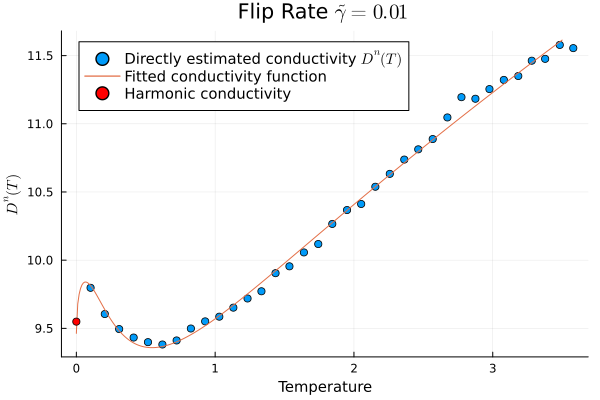}
	\includegraphics[width = 0.49\textwidth]{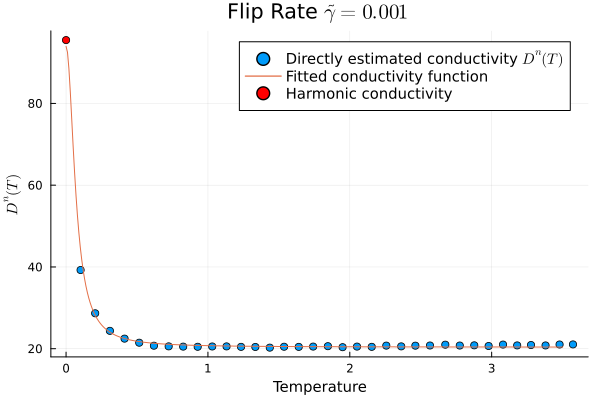}
	\includegraphics[width = 0.49\textwidth]{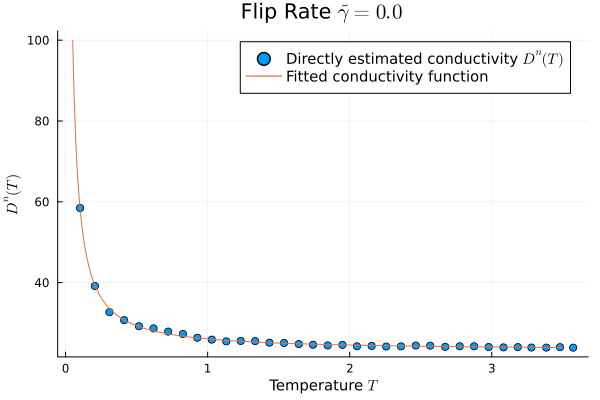}
	\caption{Thermal conductivity computed for the simulated temperatures (blue points) and the functions we fitted to these values (orange lines). We plot with a red point at \(T = 0\) the conductivity in the harmonic case, which is temperature independent.}
	\label{fig:cond_fit}
\end{figure}
\begin{itemize}
	\item For \(\widetilde{\gamma} \in \left\{1.0, 0.316, 0.1, 0.0316\right\}\) we fit the function
	\[\widehat{D}(T) = c_1 + c_2 T + c_3 \sqrt{T}.\]
	
	\item For \(\widetilde{\gamma} \in \left\{0.01, 0.00316\right\}\) we fit the function
	\[\widehat{D}(T) = c_1 + c_2 T + c_3 \sqrt{T} + \frac{c_4}{k + T^{1.5}}.\]
	
	\item For \(\widetilde{\gamma} = 0.001\) we fit the function
	\[\widehat{D}(T) = c_1 + \frac{c_2}{k + T^{2}}.\]
	
	\item For \(\widetilde{\gamma} = 0.0\) we fit the function
	\[\widehat{D}(T) = c_1 + \frac{c_2}{T^{1.1}}\]
\end{itemize}
To approximate the function \( T \mapsto \mathbb{W}^n\left(T, f_0, \theta\right) \), we use linear interpolation between the temperature values at which \(\mathbb{W}^n\) was computed via \eqref{eq:split_W_integrals}. Figure~\ref{fig:work_fit} shows four examples of the resulting curves. 

\paragraph{Temperatures used for comparison of bulk energy flux and rate of work}
We report in Table~\ref{tab:flip1} temperature values used for estimated rate of work in Figure~\ref{fig:emp_flux_v_work_flip1}. In Table~\ref{tab:flip0.1}, we have the temperatures used in Figure~\ref{fig:emp_flux_v_work_flip01} and in Table~\ref{tab:flip0}, we have those used in Figure~\ref{fig:emp_flux_v_work_noflip}.
\begin{table}[h]
	\centering
	\begin{tabular}{|cc||c|c|c|}
		\hline
		&& \multicolumn{3}{c|}{Forcing strength \(f_0\)}\\
		&& \multicolumn{1}{c}{0.5} & \multicolumn{1}{c}{1.0} & \multicolumn{1}{c|}{2.0}\\
		\hline\hline
		\parbox[t]{2mm}{\multirow{10}{*}{\rotatebox[origin=c]{90}{Forcing frequency \(\nu\)}}} & 0.15  & 0.375 & 0.889 & 2.114\\
		& 0.16  & 0.377 & 0.960 & 2.281\\
		& 0.175 & 0.405 & 1.012 & 2.263\\
		& 0.2   & 0.432 & 1.051 & 2.398\\
		& 0.225 & 0.440 & 1.057 & 2.557\\
		& 0.25  & 0.430 & 1.075 & 2.644\\
		& 0.275 & 0.446 & 1.115 & 2.624\\
		& 0.3   & 0.414 & 1.115 & 2.687\\
		& 0.325 & 0.412 & 1.061 & 2.636\\
		& 0.35  & 0.414 & 1.052 & 2.659\\ \hline
	\end{tabular}
	\caption{Temperature observed at forced site for the chain with flip rate \(\widetilde{\gamma} = 1\). }\label{tab:flip1}
\end{table}

\begin{table}[h]
	\centering
	\begin{tabular}{|cc||c|c|c|c|c|c|}
		\hline
		&& \multicolumn{6}{c|}{Forcing strength \(f_0\)}\\
		&& \multicolumn{1}{c}{0.5} & \multicolumn{1}{c}{1.0} & \multicolumn{1}{c}{2.0} & \multicolumn{1}{c}{3.0} & \multicolumn{1}{c}{4.0} & \multicolumn{1}{c|}{5.0}\\
		\hline\hline
		\parbox[t]{2mm}{\multirow{10}{*}{\rotatebox[origin=c]{90}{Forcing frequency \(\nu\)}}}& 0.15 & 0.192 & 0.404 & 0.981 & 1.671 & 2.451 & 3.221\\
		& 0.16  & 0.241 & 0.544 & 1.358 & 2.314 & 3.305 & 4.425\\
		& 0.175 & 0.256 & 0.583 & 1.463 & 2.470 & 3.515 & 4.678\\
		& 0.2   & 0.227 & 0.511 & 1.265 & 2.154 & 3.131 & 4.122\\
		& 0.225 & 0.209 & 0.461 & 1.139 & 1.939 & 2.815 & 3.751\\
		& 0.25  & 0.195 & 0.422 & 1.035 & 1.796 & 2.630 & 3.530\\
		& 0.275 & 0.187 & 0.391 & 0.984 & 1.684 & 2.483 & 3.314\\
		& 0.3   & 0.176 & 0.363 & 0.920 & 1.594 & 2.353 & 3.165\\
		& 0.325 & 0.164 & 0.332 & 0.860 & 1.504 & 2.245 & 3.014\\
		& 0.35  & 0.151 & 0.299 & 0.792 & 1.420 & 2.123 & 2.868\\
		\hline\hline
	\end{tabular}
	\caption{Temperatures observed at forced site for the chain with flip rate \(\widetilde{\gamma} = 0.1\). }\label{tab:flip0.1}
\end{table}

\begin{table}[h]
	\centering
	\begin{tabular}{|cc||c|c|c|c|c|c|}
		\hline
		&& \multicolumn{6}{c|}{Forcing strength \(f_0\)}\\
		&& \multicolumn{1}{c}{0.5} & \multicolumn{1}{c}{1.0} & \multicolumn{1}{c}{2.0} & \multicolumn{1}{c}{3.0} & \multicolumn{1}{c}{4.0} & \multicolumn{1}{c|}{5.0}\\
		\hline\hline
		\parbox[t]{2mm}{\multirow{10}{*}{\rotatebox[origin=c]{90}{Forcing frequency \(\nu\)}}}& 0.15  & 0.101 & 0.101 & 0.105 & 0.112 & 0.121 & 0.132\\
		& 0.16  & 0.157 & 0.221 & 1.250 & 1.197 & 1.325 & 1.466\\
		& 0.175 & 0.102 & 0.105 & 0.118 & 0.133 & 0.151 & 0.176\\
		& 0.2   & 0.101 & 0.104 & 0.118 & 0.143 & 0.186 & 0.253\\
		& 0.225 & 0.101 & 0.105 & 0.120 & 0.152 & 0.214 & 0.308\\
		& 0.25  & 0.101 & 0.104 & 0.119 & 0.152 & 0.216 & 0.317\\
		& 0.275 & 0.101 & 0.104 & 0.119 & 0.149 & 0.210 & 0.312\\
		& 0.3   & 0.101 & 0.104 & 0.117 & 0.144 & 0.196 & 0.293\\
		& 0.325 & 0.101 & 0.104 & 0.114 & 0.136 & 0.178 & 0.262\\
		& 0.35  & 0.101 & 0.103 & 0.111 & 0.127 & 0.158 & 0.215\\
		\hline\hline
	\end{tabular}
	\caption{Temperatures observed at forced site for the chain without flip.}\label{tab:flip0}
\end{table}

\subsection{Analytic expressions of tail integrals}\label{sec:tail_integrals}
In this subsection, we give analytic expressions of the second integral in \eqref{eq:split_W_integrals}--hereafter referred to as the tail integral--expressed in terms of the forcing and fitting parameters. These expressions are obtained by replacing the momentum autocorrelation with the fitted function describing its tail behavior. 
We start by considering the following general expression of the tail integral
\[
\int_{t_0}^{\infty} \cos(2\pi \nu t) \,\mathbb{E}_{\mu^n_T}[p_0(0)\,p_0(t)] \dd t \approx K \int_{t_0}^{\infty} \cos(2\pi \nu t)\, \hat{C}(t)\, \dd t, \qquad i = 1,\dots,4.
\]
The generic function
\[ \hat{C}(t) = K \cos\left[a\,(2\pi \nu^\star t + \phi)\right] \mathrm{e}^{-\lambda t} t^{-\alpha},\]
with \(a\in\{0,1\}\) \(\alpha \in\{0, 1/2\}\) and \(\lambda\ge0\), represents all \( \hat{C}_{p,i}(t; K,\lambda,\nu^\star, \phi),\, i = 1,\dots,4\) given in~\eqref{eq:Cp1}-\eqref{eq:Cp4}.

We define the integral
\begin{equation}\label{eq:Ip_def}
I^{p}_{a, \alpha, \lambda}(t_0) = K \int_{t_0}^{\infty} 
\cos(2\pi \nu t) \cos\left[a\,(2\pi \nu^\star t + \phi)\right] 
\mathrm{e}^{-\lambda t} t^{-\alpha}	\dd t,
\end{equation}
and introduce
\[ \Omega_{\pm} = \lambda + i 2\pi(a\nu^\star \pm \nu).
\]
Since 
\(\cos(\beta) \, cos(\gamma) = \frac 12 [cos(\beta+\gamma)] + \cos(\beta-\gamma)] 
= \frac 12 \left[\mathfrak{R}\left(\mathrm{e}^{i (\beta + \gamma)} + \mathrm{e}^{-i (\beta + \gamma)}\right) \right],\) 
we can rewrite \eqref{eq:Ip_def} as
\[ I^{p}_{a, \alpha, \lambda}(t_0) = 
\frac K2  \mathfrak{R}\left(\mathrm{e}^{-ia\phi} \int_{t_0}^{+\infty} t^{-\alpha} (\mathrm{e}^{-\Omega_+ t} +\mathrm{e}^{-\Omega_- t}) \dd t\right).
\]		
The two integrals can be rewritten as
\[
\int_{t_0}^{\infty} \mathrm{e}^{-\Omega_\pm t} t^{-\alpha} \dd t = \int_{\Omega_\pm t_0}^{\infty} \mathrm{e}^{-u} u^{-\alpha} \dd u = \Omega_\pm^{\alpha -1} \,\Gamma(1-\alpha, \Omega_\pm t_0),
\]
where we made the change of variable \(u = \Omega_\pm t\) and $\Gamma(s,z)$ is the upper incomplete Gamma function
\[
\Gamma(s,z) = \int_z^\infty \mathrm{e}^{-u} u^{s-1} \dd u,
\]
so the general expression for the approximation of the tail integral is 
\begin{equation}\label{eq:Ip_expr}
\begin{aligned} 
I^{p}_{a, \alpha, \lambda}(t_0) = \frac K2 \left[\right.
&\cos(a\phi) \, \mathfrak{R}\left( 
\Omega_+^{\alpha-1} \Gamma(1-\alpha, \Omega_+ t_0)+
\Omega_-^{\alpha-1} \Gamma(1-\alpha, \Omega_- t_0)
 \right)\\
+&\left.  \sin(a\phi) \, \mathfrak{I}\left( 
\Omega_+^{\alpha-1} \Gamma(1-\alpha, \Omega_+ t_0)+
\Omega_-^{\alpha-1} \Gamma(1-\alpha, \Omega_- t_0)
\right)
\right].
\end{aligned}
\end{equation}
For \(\alpha\in\{0.1/2\}\), the upper incomplete Gamma function has the following forms
\begin{equation}\label{eq:Gamma}
\Gamma(1-\alpha, \Omega_\pm t_0) 
= \left\{
\begin{aligned}
	&\, \mathrm{e}^{-\Omega_\pm t_0},	& \quad \text{for}\quad \alpha=0, \\[0.5em]
	&\, \sqrt{\pi}\, \mathrm{erfc}\left(\sqrt{\Omega_\pm t_0}\right), &\quad \text{for}\quad \alpha=\frac 12,
\end{aligned}
\right.
\end{equation}
where \(\mathrm{erfc}(z_0) = \frac 2{\sqrt\pi}\int_z^{+\infty} \mathrm{e}^{-z^2} \dd z\) is the complementary error function.

\paragraph{Computing $I^p_{a,0,\lambda}$, \(a\in\{0,1\}\).} In this case \(|\Omega_\pm|^2 = \lambda^2 + (2\pi(a\nu^\star\pm \nu))^2\). We have the following:
\[
I^p_{a,0,\lambda}(t_0) = \frac K2 
\left[
\cos(a\phi)\,
\mathfrak{R}\left( 
\Omega_+^{-1} \mathrm{e}^{-\Omega_+ t_0} + \Omega_-^{-1} \mathrm{e}^{-\Omega_- t_0}\right) 
+ \sin(a\phi)\, 
\mathfrak{I}\left( 
\Omega_+^{-1} \mathrm{e}^{-\Omega_+ t_0} + \Omega_-^{-1} \mathrm{e}^{-\Omega_- t_0}\right)\right].
\]
Considering that
\begin{gather*}
	\mathfrak{R}\left(\Omega_\pm^{-1} \mathrm{e}^{-\Omega_\pm t_0}\right) = \frac{\mathrm{e}^{-\lambda t_0}}{|\Omega_\pm|^2} \mathfrak{R}\left(\overline\Omega_\pm \mathrm{e}^{i 2\pi(a\nu^\star \pm \nu)t_0}\right)\\ 
	= \frac{\mathrm{e}^{-\lambda t_0}}{|\Omega_\pm|^2} \left(\lambda \cos(2\pi(a\nu^\star \pm \nu)t_0)
	+ 2\pi(a\nu^\star \pm \nu) \sin(2\pi(a\nu^\star \pm \nu)t_0) \right),
\end{gather*}
and
\begin{gather*}
	\mathfrak{I}\left(\Omega_\pm^{-1} \mathrm{e}^{-\Omega_\pm t_0}\right) =
	\frac{\mathrm{e}^{-\lambda t_0}}{|\Omega_\pm|^2}
	\mathfrak{I}\left(\overline\Omega_\pm \mathrm{e}^{i 2\pi(a\nu^\star \pm \nu)t_0}\right)\\
	= \frac{\mathrm{e}^{-\lambda t_0}}{|\Omega_\pm|^2}
	\left(\lambda \sin(2\pi(a\nu^\star \pm \nu)t_0) - 2\pi(a\nu^\star \pm \nu) \cos(2\pi(a\nu^\star \pm \nu)t_0) \right),
\end{gather*}
the final expression is
\begin{equation}\label{eq:Ip1-2_expr}
	\begin{aligned}
		I^p_{a,0,\lambda}(t_0) =  \frac K2 \mathrm{e}^{-\lambda t_0}& 
		\left[\cos(a\phi)
		\left(\frac{\lambda \cos(2\pi(a\nu^\star + \nu)t_0)
			+ 2\pi(a\nu^\star +\nu) \sin(2\pi(a\nu^\star + \nu)t_0)}{\lambda^2 + (2\pi(a\nu^\star+\nu))^2}\right.\right.\\
		&\left.
		\phantom{\cos(a\phi)}
		+\frac{\lambda \cos(2\pi(a\nu^\star - \nu)t_0)
			+ 2\pi(a\nu^\star -\nu) \sin(2\pi(a\nu^\star - \nu)t_0)}{\lambda^2 + (2\pi(a\nu^\star-\nu))^2}
		\right)\\
		&\left.+ \sin(a\phi)
		\left(\frac{\lambda \sin(2\pi(a\nu^\star + \nu)t_0)
			- 2\pi(a\nu^\star +\nu) \cos(2\pi(a\nu^\star + \nu)t_0)}{\lambda^2 + (2\pi(a\nu^\star+\nu))^2}\right.\right.\\
		&\left.\left.
		\phantom{\cos(a\phi)}
		+\frac{\lambda \sin(2\pi(a\nu^\star - \nu)t_0)
			- 2\pi(a\nu^\star -\nu) \cos(2\pi(a\nu^\star - \nu)t_0)}{\lambda^2 + (2\pi(a\nu^\star-\nu))^2}
		\right)
		\right].
	\end{aligned}
\end{equation}

Setting \(a=0\) in the previous expression (so that \(\Omega_+ = \overline\Omega_-\), the bar denoting the complex conjugation, and \(|\Omega_\pm|^2 = \lambda^2 + (2\pi \nu)^2\)), one obtains
\begin{equation}\label{eq:Ip1_expr}
	I^p_{0,0,\lambda}(t_0) = \frac{K\,\mathrm{e}^{-\lambda t_0}}{\lambda^2 + (2\pi\nu)^2} \, (\lambda \cos(2\pi\nu t_0) - 2\pi\nu \sin(2\pi\nu t_0)),
\end{equation}
while the tail integral \(I^p_{1,0,\lambda}(t_0)\) can be easily deduced from~\eqref{eq:Ip1-2_expr} by setting \(a=1\).

\paragraph{Computing $I^p_{1,1/2,\lambda}$.} Once again \(|\Omega_\pm|^2 = \lambda^2 + (2\pi(\nu^\star\pm\nu))^2\). We have the following:
\[
\begin{aligned}
I^p_{1,1/2,\lambda}(t_0) 
=  \frac{K\sqrt{\pi}}2 
&\left[\cos\phi\,\mathfrak{R}\left( 
\Omega_+^{-1/2} \mathrm{erfc}(\sqrt{\Omega_+ t_0}) + \Omega_-^{-1/2} \mathrm{erfc}(\sqrt{\Omega_- t_0})\right) \right.\\
& \left.+ \sin\phi\, \mathfrak{I}\left( 
\Omega_+^{-1/2} \mathrm{erfc}(\sqrt{\Omega_+ t_0}) + \Omega_-^{-1/2} \mathrm{erfc}(\sqrt{\Omega_- t_0})\right)\right]
\\
= \frac{K\sqrt{\pi}}{2 |\Omega_+|}
&\left[
\cos\phi\,
\left(\mathfrak{R}\left(\overline\Omega_+^{1/2}\right) \mathfrak{R}\left(\mathrm{erfc}(\sqrt{\Omega_+ t_0})\right) -
\mathfrak{I}\left(\overline\Omega_+^{1/2}\right) \mathfrak{I}\left(\mathrm{erfc}(\sqrt{\Omega_+ t_0})\right)\right)\right.\\
&\left.
+ \sin\phi\left(
\mathfrak{R}\left(\overline\Omega_+^{1/2}\right) 
\mathfrak{I}\left( \mathrm{erfc}(\sqrt{\Omega_+ t_0})\right)
+
\mathfrak{I}\left(\overline\Omega_+^{1/2}\right) 
\mathfrak{R}\left( \mathrm{erfc}(\sqrt{\Omega_+ t_0})\right)
\right)\right]\\
+\frac{K\sqrt{\pi}}{2 |\Omega_-|}
&\left[
\cos\phi\,
\left(\mathfrak{R}\left(\overline\Omega_-^{1/2}\right) \mathfrak{R}\left(\mathrm{erfc}(\sqrt{\Omega_- t_0})\right) -
\mathfrak{I}\left(\overline\Omega_-^{1/2}\right) \mathfrak{I}\left(\mathrm{erfc}(\sqrt{\Omega_- t_0})\right)\right)\right.\\
&\left.
+ \sin\phi\left(
\mathfrak{R}\left(\overline\Omega_-^{1/2}\right) 
\mathfrak{I}\left( \mathrm{erfc}(\sqrt{\Omega_- t_0})\right)
+
\mathfrak{I}\left(\overline\Omega_-^{1/2}\right) 
\mathfrak{R}\left( \mathrm{erfc}(\sqrt{\Omega_- t_0})\right)
\right)\right]\\
\end{aligned}
\]
Considering that
\begin{gather*}
(\overline\Omega_\pm)^{1/2} = \sqrt{|\Omega_\pm|} \exp(-\frac i 2 \atan(\frac{2\pi(\nu^\star\pm\nu)}{\lambda}))
= \sqrt{|\Omega_\pm|} (\cos(\frac{2\pi(\nu^\star\pm\nu)}{\lambda})) - i \sin(\frac{2\pi(\nu^\star\pm\nu)}{\lambda})))\\
\mathfrak{R}((\overline\Omega_\pm)^{1/2}) = \sqrt{|\Omega_\pm|} \cos\left(\frac12 \atan(\frac{2\pi(\nu^\star\pm\nu)}{\lambda})\right) \\
= \sqrt{|\Omega_\pm|} \sqrt{\frac12\left(1 + \frac1{\sqrt{\frac{(2\pi(\nu^\star\pm\nu))^2}{\lambda^2}+1}}\right)} 
= \sqrt{\frac12\left(|\Omega_\pm| +\lambda\right)}
\end{gather*}
and \(\sin[2](x) = 1 - \cos[2](x)\), so
\begin{equation*}
\mathfrak{R}((\overline\Omega_\pm)^{1/2}) = 
\sqrt{\frac{|\Omega_\pm| +\lambda}2},
\qquad 
\mathfrak{I}((\overline\Omega_\pm)^{1/2}) = 
- \sqrt{\frac{|\Omega_\pm| -\lambda}2},
\end{equation*}
we finally obtain the following
\begin{equation}\label{eq:Ip3-4_expr}
\begin{aligned}
I^p_{1,1/2,\lambda}(t_0) 
= \frac{K\sqrt{\pi}}{2 |\Omega_+|}
&\left[
\cos\phi\,
\left(\sqrt{\frac{|\Omega_+| +\lambda}2}
\mathfrak{R}\left(\mathrm{erfc}(\sqrt{\Omega_+ t_0})\right)
+ 
\sqrt{\frac{|\Omega_+| -\lambda}2} \mathfrak{I}\left(\mathrm{erfc}(\sqrt{\Omega_+ t_0})\right)\right)\right.\\
&\left.
+ \sin\phi\left(
\sqrt{\frac{|\Omega_+| +\lambda}2} 
\mathfrak{I}\left(\mathrm{erfc}(\sqrt{\Omega_+ t_0})\right)
-
\sqrt{\frac{|\Omega_+| -\lambda}2}
\mathfrak{R}\left( \mathrm{erfc}(\sqrt{\Omega_+ t_0})\right)
\right)\right]\\
+\frac{K\sqrt{\pi}}{2 |\Omega_-|}
&\left[
\cos\phi\,
\left(\sqrt{\frac{|\Omega_-| +\lambda}2} \mathfrak{R}\left(\mathrm{erfc}(\sqrt{\Omega_- t_0})\right)
+
\sqrt{\frac{|\Omega_-| -\lambda}2} \mathfrak{I}\left(\mathrm{erfc}(\sqrt{\Omega_- t_0})\right)\right)\right.\\
&\left.
+ \sin\phi\left(
\sqrt{\frac{|\Omega_-| +\lambda}2}
\mathfrak{I}\left( \mathrm{erfc}(\sqrt{\Omega_- t_0})\right)
-
\sqrt{\frac{|\Omega_-| -\lambda}2} \mathfrak{R}\left( \mathrm{erfc}(\sqrt{\Omega_- t_0})\right)
\right)\right].\\
\end{aligned}
\end{equation}

\paragraph{Expression of \(I^p_{1,1/2,0}\).}
We can obtain a more explicit expression for \(I^p_{1,1/2,0}\) from ~\eqref{eq:Ip3-4_expr}, by noting that, in this case, $\Omega_\pm = i 2\pi(\nu^\star\pm\nu)$, \emph{i.e.} pure imaginary. We can thus express the error function via the Fresnel cosine and sine integrals, given respectively by
\[\C\left(y_0\right) = \int_0^{y_0}\cos\left(\frac{\pi}{2}y^2\right) \dd y, \qquad \S\left(y_0\right) = \int_0^{y_0} \sin\left(\frac{\pi}{2}y^2\right)\dd y,\]
see \cite[Section 7.3]{Abramowitz_Stegun} or \cite[Section 7.2]{NIST:DLMF}. 
Indeed, introducing $z^\pm_0 = \sqrt{i\, 2\pi(\nu^\star\pm\nu)  t_0}$, we have
\begin{gather*}
\erf(z^\pm_0) = \frac 2{\sqrt{\pi}} \int_0^{z^\pm_0} \mathrm{e}^{-z^2} \dd z 
= \frac2{\sqrt{2}} \mathrm{e}^{i\, \mathrm{sign}(\nu^\star \pm \nu)\pi/4} 
\int_0^{y^\pm_0} \mathrm{e}^{-i\, \mathrm{sign}(\nu^\star\pm\nu)\frac\pi2 y^2} \dd y\\
= \sqrt{2} \mathrm{e}^{i\, \mathrm{sign}(\nu^\star \pm \nu)\pi/4} \left[\int_0^{y^\pm_0} \cos(\frac\pi2 y^2) \dd y - i\, \mathrm{sign}(\nu^\star \pm \nu) \int_0^{y^\pm_0} \sin(\frac\pi2 y^2) \dd y\right]\\
= (1 + i\, \mathrm{sign}(\nu^\star \pm \nu)) \left( \C(y^\pm_0) - i\, \mathrm{sign}(\nu^\star \pm \nu) \S(y^\pm_0)\right)
\end{gather*}
where we performed the change of variable \(z = \sqrt{i\,\mathrm{sign}(\nu^\star \pm \nu))\frac\pi2}\, y\)  with $y\in\mathbb R$ and \(y^\pm_0 = \sqrt{\frac{2}{\pi}}  |z^\pm_0|\).
We obtain
\begin{equation}\label{eq:erfc}
\begin{aligned}
\mathrm{erfc}(\sqrt{\Omega_\pm t_0}) = & 1 - \,\erf(\sqrt{\Omega_\pm t_0)}\\ 
= & 1 - \C\left(2 \sqrt{|\nu^\star\pm\nu| t_0}\right) - \S\left(2 \sqrt{|\nu^\star\pm\nu| t_0}\right) \\
& + i\,\mathrm{sign}(\nu^\star \pm \nu) \left(\S\left(2 \sqrt{|\nu^\star\pm\nu| t_0}\right) - \C\left(2 \sqrt{|\nu^\star\pm\nu| t_0}\right)\right).
\end{aligned}
\end{equation}
Thus, from~\eqref{eq:Ip3-4_expr}
\begin{equation*}
	\begin{aligned}
		I^p_{1,1/2,0}(t_0) 
		= \frac{K\sqrt{\pi}}{\sqrt{2 |\Omega_+|}}
		&\left[
		\frac{\cos\phi}2\,
		\left(\mathfrak{R}\left(\mathrm{erfc}(\sqrt{\Omega_+ t_0})\right)
		+ 
		\mathfrak{I}\left(\mathrm{erfc}(\sqrt{\Omega_+ t_0})\right)\right)\right.\\
		&\left.
		+ \frac{\sin\phi}2\left( 
		\mathfrak{I}\left(\mathrm{erfc}(\sqrt{\Omega_+ t_0})\right)
		-
		\mathfrak{R}\left( \mathrm{erfc}(\sqrt{\Omega_+ t_0})\right)
		\right)\right]\\
		+\frac{K\sqrt{\pi}}{\sqrt{2 |\Omega_-|}}
		&\left[
		\frac{\cos\phi}2\,
		\left(\mathfrak{R}\left(\mathrm{erfc}(\sqrt{\Omega_- t_0})\right)
		+
		\mathfrak{I}\left(\mathrm{erfc}(\sqrt{\Omega_- t_0})\right)\right)\right.\\
		&\left.
		+ \frac{\sin\phi}2\,
		\left(
		\mathfrak{I}\left( \mathrm{erfc}(\sqrt{\Omega_- t_0})\right)
		-
		\mathfrak{R}\left( \mathrm{erfc}(\sqrt{\Omega_- t_0})\right)
		\right)\right],
	\end{aligned}
\end{equation*}
which, by~\eqref{eq:erfc}, gives
\begin{equation}\label{eq:Ip4_expr}
\begin{aligned}
I^p_{1,1/2,0}(t_0)
= \frac{K}{\sqrt{|\nu^\star+\nu|}}
&\left[
\frac{\cos\phi}2\,
\left(\frac12 - \C\left(2 \sqrt{|\nu^\star+\nu| t_0}\right) \right)
- \frac{\sin\phi}2\,
\left(\frac12 - \S\left(2 \sqrt{|\nu^\star+\nu| t_0}\right) \right)\right]
\\
+\frac{K}{\sqrt{|\nu^\star-\nu|}}
&\left\{
\frac{\cos\phi}4
\left[1 - (\sgn{\nu^\star-\nu} +1)\,\C\left(2 \sqrt{|\nu^\star-\nu| t_0}\right) 
\right.
\right.
\\
&
\left.
\qquad\quad\quad\; +\, (\sgn{\nu^\star-\nu} -1)\,\S\left(2 \sqrt{|\nu^\star-\nu| t_0}\right)
\right]\\
&
- \frac{\sin\phi}4
\left[1 + (\sgn{\nu^\star-\nu} - 1)\,\C\left(2 \sqrt{|\nu^\star-\nu| t_0}\right) 
\right.
\\
&
\left.
\left.
\qquad\quad\quad\; -\, (\sgn{\nu^\star-\nu} + 1)\,\S\left(2 \sqrt{|\nu^\star-\nu| t_0}\right)
\right]
\right\}.
	\end{aligned}
\end{equation}

\section{Computation of solution to limit equation}\label{sec:solving_pde}
In this appendix, we detail the numerical method we use for solving the equation \eqref{eq:limite_hydro}. We solve the fixed point problem \eqref{eq:fixed_point_problem} by discretizing the problem and iterating the discretized map. One can hope that, for an initial condition sufficiently close to \(T_{\mathrm{ss}}\), iterating the map \(\mathscr{M}\) generates a convergent sequence and that the same would hold after discretization.

To discretize the fixed point problem, let \(\mathscr{U}^K = \left\{0 < u_1 < \cdots < u_K = 1\right\}\) be a discretization mesh of~\([0, 1]\), and \(D^n(T)\) and \(\mathbb{W}^n\left(T, f_0, \theta\right)\) respectively be the numerical approximations of the thermal conductivity and the rate of work, the estimation of which was described in Appendix~\ref{sec:correlation}. For a vector \(T^k \in \mathbb{R}^{K+1}\), our discretization of \eqref{eq:M}, \(\widehat{\mathscr{M}}:\mathbb{R}^{K+1} \to \mathbb{R}^{K+1}\), is given by
\[\widehat{\mathscr{M}}\left[T^k\right]_i = \left\{\begin{aligned} & T_\ell + \mathbb{W}^n\left(T^k_{K}, f_0, \theta \right) \sum_{j = 1}^i \frac{u_j - u_{j-1}}{2}\left(\frac{1}{D^n\left(T_j\right)} + \frac{1}{D^n\left(T_{j-1}\right)}\right) \quad && i \in \left\{1, \dots, K\right\},\\
	& T_\ell && i = 0.
\end{aligned} \right.\]
For plotting solutions to \eqref{eq:limite_hydro} in Section~\ref{sec:numerics}, we used a uniform mesh with \(K = 2000\) discretization points. We initialize the iteration \(T^{k+1} = \widehat{\mathscr{M}}\left[T^k\right]\) with \(T_i^0 = T_\ell\) for all \(i \in \left\{0, 1, \dots, K\right\}\) and stop the iteration when
\(\max_{i \in \left\{1, \dots, K\right\}}\left|T_i^k - T_i^{k-1}\right| < 10^{-10}\).

\section{Numerical Verification of the Sufficient Condition for the Uniqueness of the Solution to \eqref{eq:limite_hydro}}\label{sec:numerical_verification_strict_increasing}
\begin{figure}[h]
\centering
       \includegraphics[width = 0.49\textwidth]{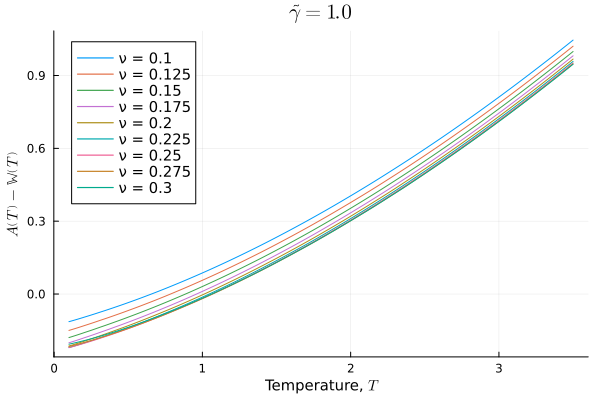}
       \includegraphics[width = 0.49\textwidth]{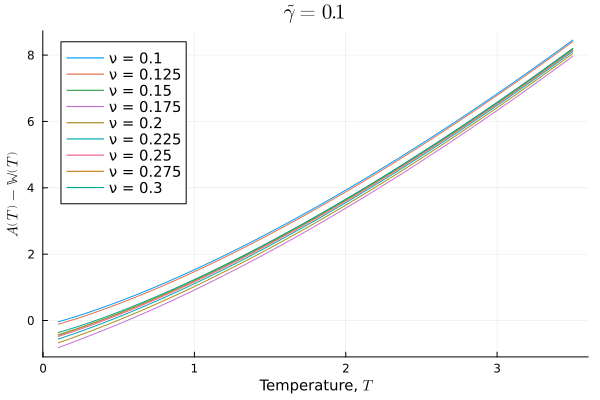}
       \includegraphics[width = 0.49\textwidth]{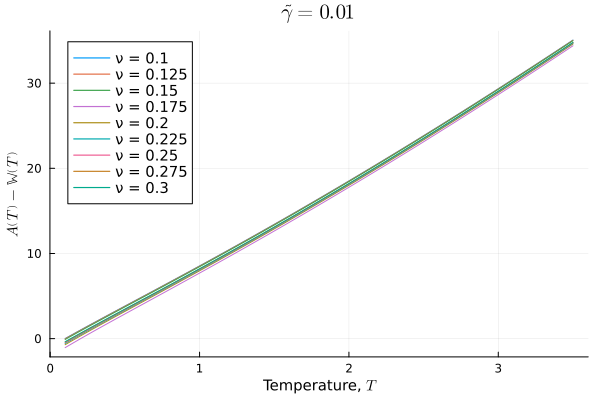}
       \includegraphics[width = 0.49\textwidth]{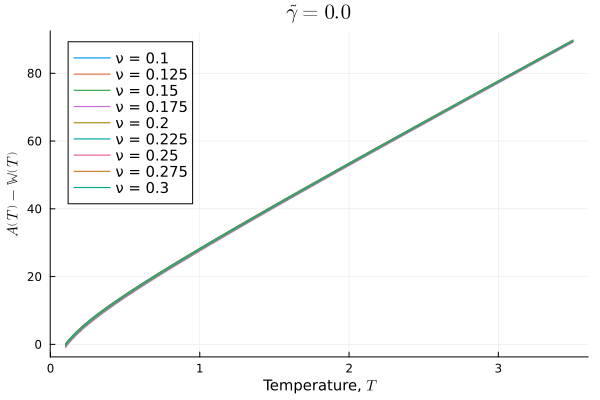}
       \caption{Numerical estimation of the function \(T \mapsto A\left(T\right) - \mathbb{W}\left(T, \mathdutchcal{F}, \theta\right)\) for several values of forcing frequencies and flip rates (including without flip).}\label{fig:estimated_A-W}
\end{figure}
We plot in Figure~\ref{fig:estimated_A-W} the numerically estimated values of the function \(T \mapsto A\left(T\right) - \mathbb{W}\left(T, \mathdutchcal{F}, \theta\right)\) for several values of the forcing frequencies and the flip rates. For all the parameters we simulated, our numerical estimate of this function is strictly increasing over the range of temperature we tested.

\section{Additional Plots}\label{sec:additional_plots}
We collect together here for completeness additional plots not included in the main article and the earlier appendices due to their redundancy.

\begin{figure}[H]
	\centering
	\includegraphics[width = 0.49\textwidth]{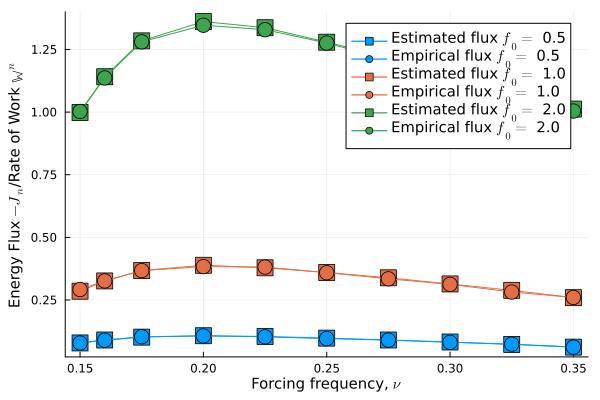}
	\includegraphics[width = 0.49\textwidth]{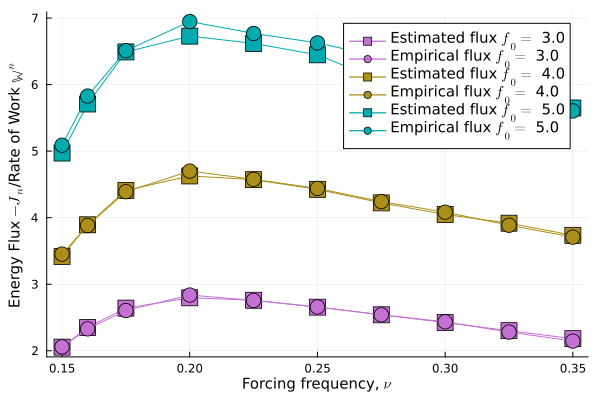}
	\caption{Empirical negative flux observed in simulations with flip rate \(\widetilde{\gamma} = 0.316\) (circles), and work rate estimated using \eqref{eq:finite_size_W} (squares), both plotted against frequency within the harmonic band, using the same forcing parameters and evaluated at the temperature of the forced atom site.}
	\label{fig:emp_flux_v_work_flip0316}
\end{figure}

\begin{table}[H]
	\centering
	\begin{tabular}{|cc||c|c|c|c|c|c|}
		\hline
		&& \multicolumn{6}{c|}{Forcing strength \(f_0\)}\\
		&& \multicolumn{1}{c}{0.5} & \multicolumn{1}{c}{1.0} & \multicolumn{1}{c}{2.0} & \multicolumn{1}{c}{3.0} & \multicolumn{1}{c}{4.0} & \multicolumn{1}{c|}{5.0}\\
		\hline\hline
		\parbox[t]{2mm}{\multirow{10}{*}{\rotatebox[origin=c]{90}{Forcing frequency \(\nu\)}}}& 0.15 & 0.192 & 0.404 & 0.981 & 1.671 & 2.451 & 3.221\\
		& 0.16  & 0.241 & 0.544 & 1.358 & 2.314 & 3.305 & 4.425\\
		& 0.175 & 0.256 & 0.583 & 1.463 & 2.470 & 3.515 & 4.678\\
		& 0.2   & 0.227 & 0.511 & 1.265 & 2.154 & 3.131 & 4.122\\
		& 0.225 & 0.209 & 0.461 & 1.139 & 1.939 & 2.815 & 3.751\\
		& 0.25  & 0.195 & 0.422 & 1.035 & 1.796 & 2.630 & 3.530\\
		& 0.275 & 0.187 & 0.391 & 0.984 & 1.684 & 2.483 & 3.314\\
		& 0.3   & 0.176 & 0.363 & 0.920 & 1.594 & 2.353 & 3.165\\
		& 0.325 & 0.164 & 0.332 & 0.860 & 1.504 & 2.245 & 3.014\\
		& 0.35  & 0.151 & 0.299 & 0.792 & 1.420 & 2.123 & 2.868\\
		\hline\hline
	\end{tabular}
	\caption{Temperature observed at forced site for the chain with flip rate \(\widetilde{\gamma} = 0.316\) and used to estimate rate of work in Figure~\ref{fig:emp_flux_v_work_flip0316}. }\label{tab:flip0316}
\end{table}

\begin{figure}[H]
	\centering
	\includegraphics[width = 0.6\textwidth]{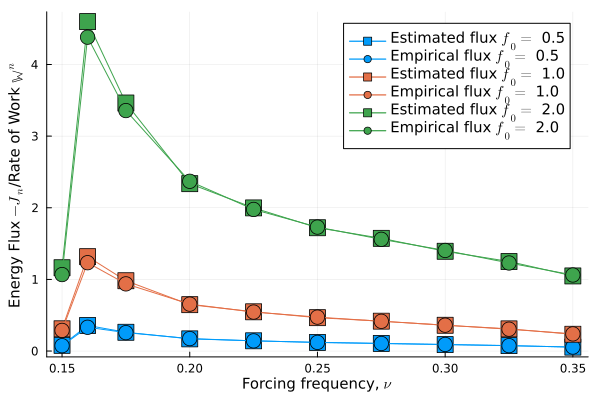}
	\caption{Empirical negative flux observed in simulations with flip rate \(\widetilde{\gamma} = 0.0316\) (circles), and work rate estimated using \eqref{eq:finite_size_W} (squares), both plotted against frequency within the harmonic band, using the same forcing parameters and evaluated at the temperature of the forced atom site.}
	\label{fig:emp_flux_v_work_flip00316}
\end{figure}

\begin{table}[H]
	\centering
	\begin{tabular}{|cc||c|c|c|}
		\hline
		&& \multicolumn{3}{c|}{Forcing strength \(f_0\)}\\
		&& \multicolumn{1}{c}{0.5} & \multicolumn{1}{c}{1.0} & \multicolumn{1}{c|}{2.0}\\
		\hline\hline
		\parbox[t]{2mm}{\multirow{10}{*}{\rotatebox[origin=c]{90}{Forcing frequency \(\nu\)}}} & 0.15  & 0.375 & 0.889 & 2.114\\
		& 0.16  & 0.377 & 0.960 & 2.281\\
		& 0.175 & 0.405 & 1.012 & 2.263\\
		& 0.2   & 0.432 & 1.051 & 2.398\\
		& 0.225 & 0.440 & 1.057 & 2.557\\
		& 0.25  & 0.430 & 1.075 & 2.644\\
		& 0.275 & 0.446 & 1.115 & 2.624\\
		& 0.3   & 0.414 & 1.115 & 2.687\\
		& 0.325 & 0.412 & 1.061 & 2.636\\
		& 0.35  & 0.414 & 1.052 & 2.659\\ \hline
	\end{tabular}
	\caption{Temperature observed at forced site for the chain with flip rate \(\widetilde{\gamma} = 0.0316\) and used to estimate rate of work in Figure~\ref{fig:emp_flux_v_work_flip00316}. }\label{tab:flip00316}
\end{table}

\begin{figure}[H]
	\centering
	\includegraphics[width = 0.6\textwidth]{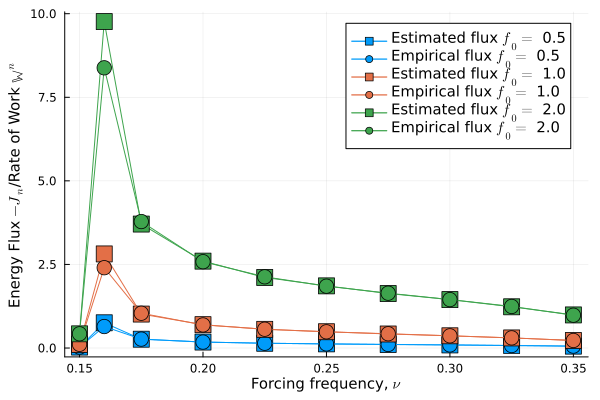}
	\caption{Empirical negative flux observed in simulations with flip rate \(\widetilde{\gamma} = 0.01\) (circles), and work rate estimated using \eqref{eq:finite_size_W} (squares), both plotted against frequency within the harmonic band, using the same forcing parameters and evaluated at the temperature of the forced atom site.}
	\label{fig:emp_flux_v_work_flip001}
\end{figure}

\begin{table}[H]
	\centering
	\begin{tabular}{|cc||c|c|c|}
		\hline
		&& \multicolumn{3}{c|}{Forcing strength \(f_0\)}\\
		&& \multicolumn{1}{c}{0.5} & \multicolumn{1}{c}{1.0} & \multicolumn{1}{c|}{2.0}\\
		\hline\hline
		\parbox[t]{2mm}{\multirow{10}{*}{\rotatebox[origin=c]{90}{Forcing frequency \(\nu\)}}} & 0.15  & 0.375 & 0.889 & 2.114\\
		& 0.16  & 0.377 & 0.960 & 2.281\\
		& 0.175 & 0.405 & 1.012 & 2.263\\
		& 0.2   & 0.432 & 1.051 & 2.398\\
		& 0.225 & 0.440 & 1.057 & 2.557\\
		& 0.25  & 0.430 & 1.075 & 2.644\\
		& 0.275 & 0.446 & 1.115 & 2.624\\
		& 0.3   & 0.414 & 1.115 & 2.687\\
		& 0.325 & 0.412 & 1.061 & 2.636\\
		& 0.35  & 0.414 & 1.052 & 2.659\\ \hline
	\end{tabular}
	\caption{Temperature observed at forced site for the chain with flip rate \(\widetilde{\gamma} = 0.01\) and used to estimate rate of work in Figure~\ref{fig:emp_flux_v_work_flip001}. }\label{tab:flip001}
\end{table}

\begin{figure}[H]
	\centering
	\includegraphics[width = 0.6\textwidth]{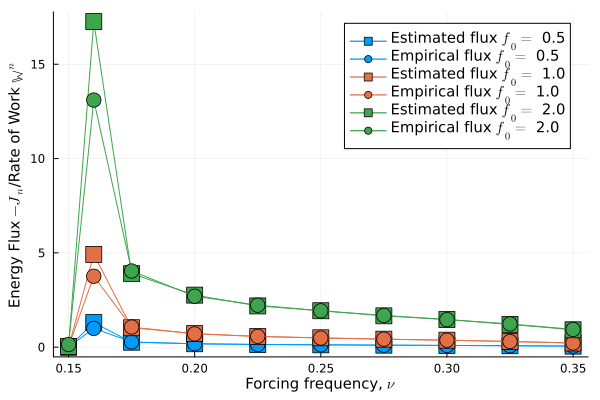}
	\caption{Empirical negative flux observed in simulations with flip rate \(\widetilde{\gamma} = 0.00316\) (circles), and work rate estimated using \eqref{eq:finite_size_W} (squares), both plotted against frequency within the harmonic band, using the same forcing parameters and evaluated at the temperature of the forced atom site.}
	\label{fig:emp_flux_v_work_flip000316}
\end{figure}

\begin{table}[H]
	\centering
	\begin{tabular}{|cc||c|c|c|}
		\hline
		&& \multicolumn{3}{c|}{Forcing strength \(f_0\)}\\
		&& \multicolumn{1}{c}{0.5} & \multicolumn{1}{c}{1.0} & \multicolumn{1}{c|}{2.0}\\
		\hline\hline
		\parbox[t]{2mm}{\multirow{10}{*}{\rotatebox[origin=c]{90}{Forcing frequency \(\nu\)}}} & 0.15  & 0.375 & 0.889 & 2.114\\
		& 0.16  & 0.377 & 0.960 & 2.281\\
		& 0.175 & 0.405 & 1.012 & 2.263\\
		& 0.2   & 0.432 & 1.051 & 2.398\\
		& 0.225 & 0.440 & 1.057 & 2.557\\
		& 0.25  & 0.430 & 1.075 & 2.644\\
		& 0.275 & 0.446 & 1.115 & 2.624\\
		& 0.3   & 0.414 & 1.115 & 2.687\\
		& 0.325 & 0.412 & 1.061 & 2.636\\
		& 0.35  & 0.414 & 1.052 & 2.659\\ \hline
	\end{tabular}
	\caption{Temperature observed at forced site for the chain with flip rate \(\widetilde{\gamma} = 0.00316\) and used to estimate rate of work in Figure~\ref{fig:emp_flux_v_work_flip000316}. }\label{tab:flip000316}
\end{table}

\begin{figure}[H]
	\centering
	\includegraphics[width = 0.6\textwidth]{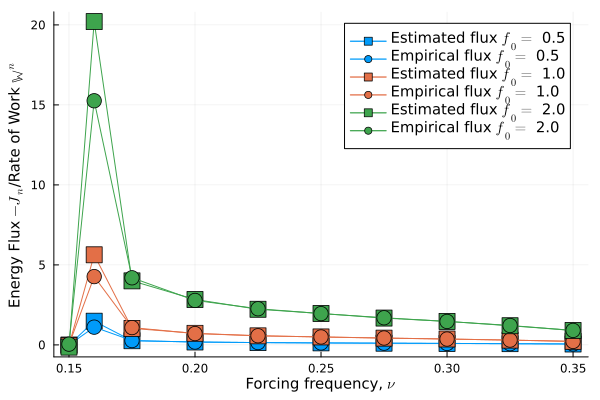}
	\caption{Empirical negative flux observed in simulations with flip rate \(\widetilde{\gamma} = 0.001\) (circles), and work rate estimated using \eqref{eq:finite_size_W} (squares), both plotted against frequency within the harmonic band, using the same forcing parameters and evaluated at the temperature of the forced atom site.}
	\label{fig:emp_flux_v_work_flip0001}
\end{figure}

\begin{table}[H]
	\centering
	\begin{tabular}{|cc||c|c|c|}
		\hline
		&& \multicolumn{3}{c|}{Forcing strength \(f_0\)}\\
		&& \multicolumn{1}{c}{0.5} & \multicolumn{1}{c}{1.0} & \multicolumn{1}{c|}{2.0}\\
		\hline\hline
		\parbox[t]{2mm}{\multirow{10}{*}{\rotatebox[origin=c]{90}{Forcing frequency \(\nu\)}}} & 0.15  & 0.375 & 0.889 & 2.114\\
		& 0.16  & 0.377 & 0.960 & 2.281\\
		& 0.175 & 0.405 & 1.012 & 2.263\\
		& 0.2   & 0.432 & 1.051 & 2.398\\
		& 0.225 & 0.440 & 1.057 & 2.557\\
		& 0.25  & 0.430 & 1.075 & 2.644\\
		& 0.275 & 0.446 & 1.115 & 2.624\\
		& 0.3   & 0.414 & 1.115 & 2.687\\
		& 0.325 & 0.412 & 1.061 & 2.636\\
		& 0.35  & 0.414 & 1.052 & 2.659\\ \hline
	\end{tabular}
	\caption{Temperature observed at forced site for the chain with flip rate \(\widetilde{\gamma} = 0.001\) and used to estimate rate of work in Figure~\ref{fig:emp_flux_v_work_flip0001}. }\label{tab:flip0001}
\end{table}

\begin{figure}[H]
	\centering
	\includegraphics[width = 0.49\textwidth]{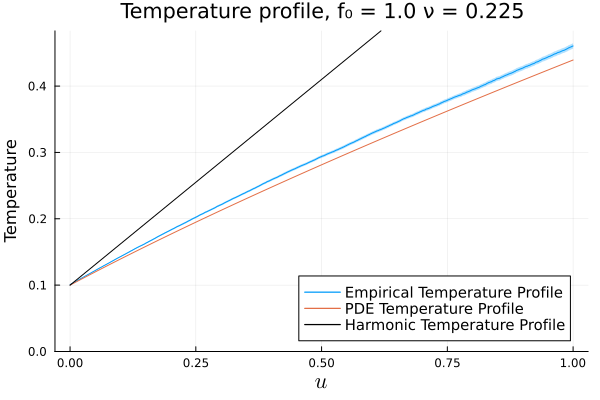}
	\includegraphics[width = 0.49\textwidth]{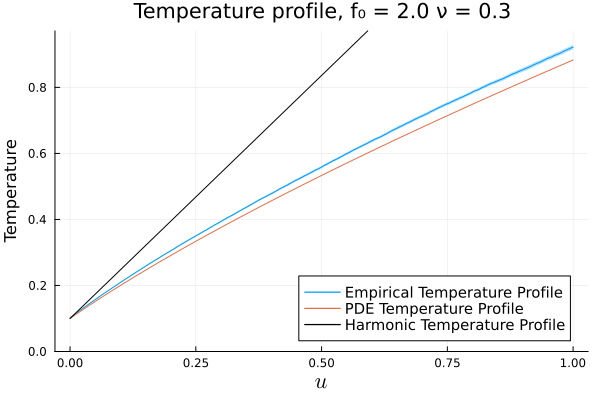}
	\caption{Temperature profile obtained by solving the PDE \eqref{eq:limite_hydro}, compared to the empirical profile computed under the same forcing parameters and flip rate \(\widetilde{\gamma} = 0.1\). Error bars around the empirical profile are shown as a blue shaded region. The harmonic temperature profile for the same forcing parameters is plotted in black.}
	\label{fig:temp_profile_flip01}
\end{figure}

\begin{figure}[H]
	\centering
	\includegraphics[width = 0.49\textwidth]{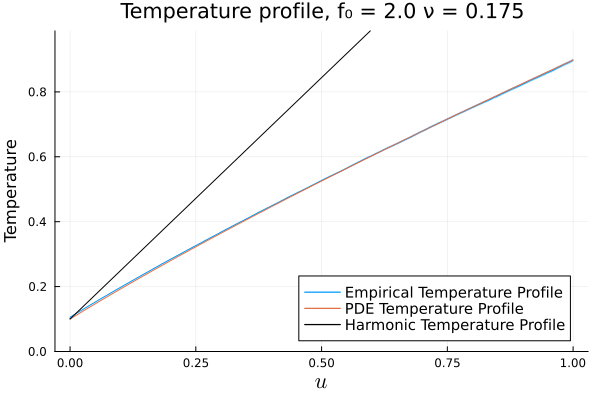}
	\includegraphics[width = 0.49\textwidth]{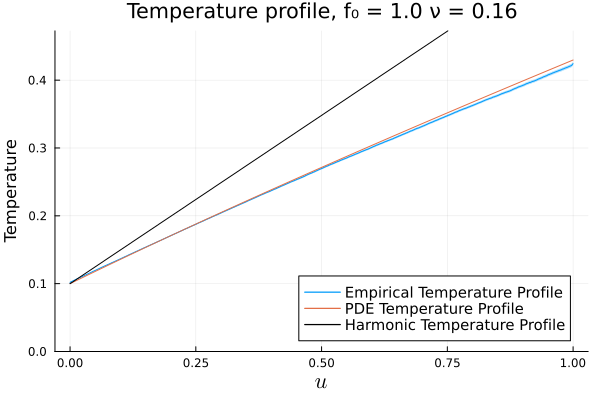}
	\caption{Temperature profile obtained by solving the PDE \eqref{eq:limite_hydro}, compared to the empirical profile computed under the same forcing parameters and flip rate \(\widetilde{\gamma} = 0.0316\). Error bars around the empirical profile are shown as a blue shaded region. The harmonic temperature profile for the same forcing parameters is plotted in black.}
	\label{fig:temp_profile_flip00316}
\end{figure}

\begin{figure}[H]
	\centering
	\includegraphics[width = 0.49\textwidth]{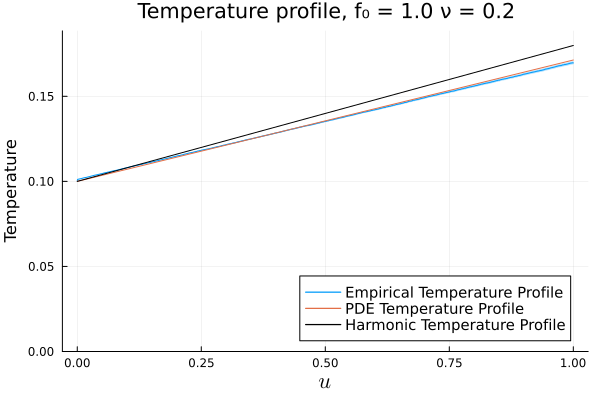}
	\includegraphics[width = 0.49\textwidth]{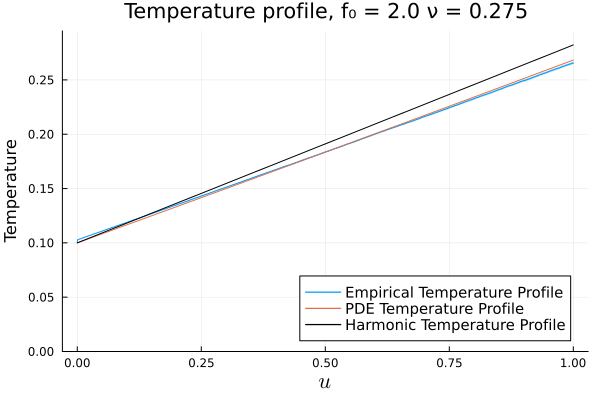}
	\caption{Temperature profile obtained by solving the PDE \eqref{eq:limite_hydro}, compared to the empirical profile computed under the same forcing parameters and flip rate \(\widetilde{\gamma} = 0.01\). Error bars around the empirical profile are shown as a blue shaded region. The harmonic temperature profile for the same forcing parameters is plotted in black.}
	\label{fig:temp_profile_flip001}
\end{figure}

\begin{figure}[H]
	\centering
	\includegraphics[width = 0.49\textwidth]{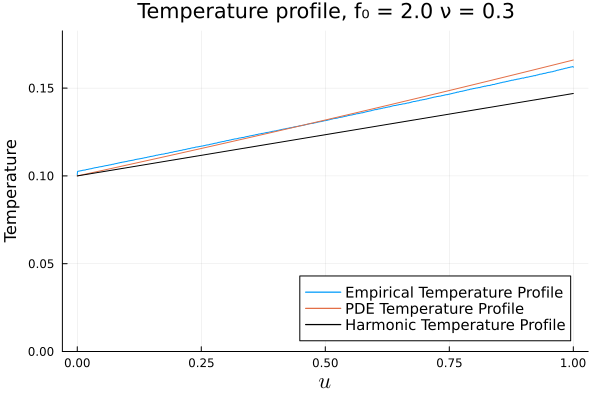}
	\includegraphics[width = 0.49\textwidth]{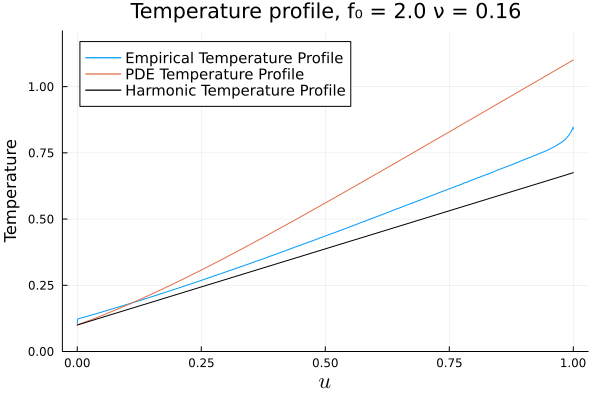}
	\caption{Temperature profile obtained by solving the PDE \eqref{eq:limite_hydro}, compared to the empirical profile computed under the same forcing parameters and flip rate \(\widetilde{\gamma} = 0.00316\). The harmonic temperature profile for the same forcing parameters is plotted in black.}
	\label{fig:temp_profile_flip000316}
\end{figure}

\begin{figure}[H]
	\centering
	\includegraphics[width = 0.49\textwidth]{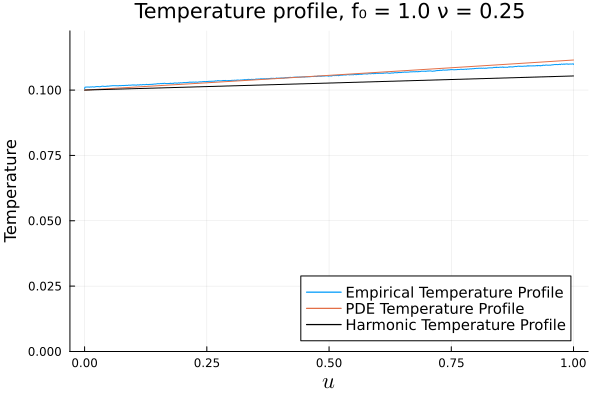}
	\includegraphics[width = 0.49\textwidth]{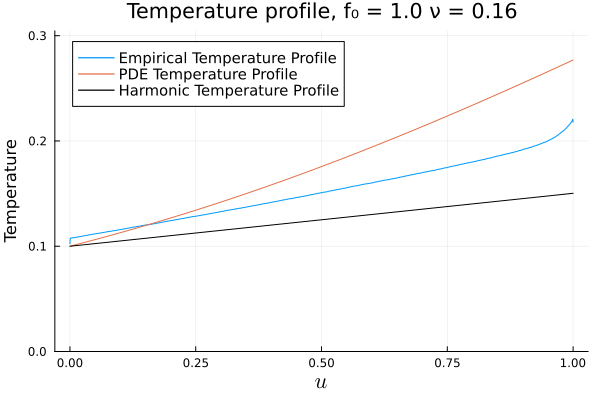}
	\caption{Temperature profile obtained by solving the PDE \eqref{eq:limite_hydro}, compared to the empirical profile computed under the same forcing parameters and flip rate \(\widetilde{\gamma} = 0.001\). The harmonic temperature profile for the same forcing parameters is plotted in black.}
	\label{fig:temp_profile_flip0001}
\end{figure}

\bigskip

\paragraph{Conflict of interest} 
The authors have no financial or non--financial conflicts of interest to declare.

\paragraph{Data Availability}
The data produced by our simulations is available at \url{https://doi.org/10.5281/zenodo.15910044}. The code for these simulations as well as the post--processing and the creation of the figures in this article is available at \url{https://github.com/shiva-darshan/beta-fput-under-periodic-forcing}.

\paragraph{Acknowledgments.}
The authors thank the two anonymous reviewers for their helpful comments and suggestions. S.D. additionally thanks Abhishek Dhar, Anupam Kundum, and Umesh Kumar for stimulating discussions at an early stage of this project. 
This work was supported by l’Agence Nationale de la Recherche under grant ANR-21-CE40-0006 (SINEQ). S.D. and G.S. also acknowledge the support of the European Research Council (ERC) under the European Union’s Horizon 2020 research and innovation programme (project EMC2, grant agreement No 810367). S.D. additionally acknowledges the support of the ERC under the European Union's Horizon Europe research and innovation programme (project UMCMC, grant agreement No 101171415).

\bibliographystyle{plain}
\bibliography{bibliography}
\end{document}